\documentclass[amssymb,prd,aps,amsmath,twocolumn,superscriptaddress,nofootinbib,preprintnumbers,10pt]{revtex4-1}
\pdfoutput=1

\usepackage[utf8]{inputenc}
\usepackage{graphicx}
\usepackage{psfrag}
\usepackage{braket}
\usepackage[usenames,dvipsnames]{color}
\usepackage{epsfig,amsmath,amssymb,fancyhdr}   
\usepackage[normalem]{ulem}
\usepackage{amsmath}
\usepackage{slashed}
\usepackage{mathtools}
\usepackage[titletoc]{appendix}
\usepackage{datetime}
\usepackage{alltt} 
\usepackage{upgreek}

\usepackage{epstopdf} 
\usepackage{listings}
\usepackage{xargs}
\usepackage{capt-of}
\usepackage{ulem}

\usepackage{array}
\newcolumntype{L}[1]{>{\raggedright\let\newline\\\arraybackslash\hspace{0pt}}m{#1}}
\newcolumntype{C}[1]{>{\centering\let\newline\\\arraybackslash\hspace{0pt}}m{#1}}
\newcolumntype{R}[1]{>{\raggedleft\let\newline\\\arraybackslash\hspace{0pt}}m{#1}}
\newcommand{\specialcell}[2][c]{%
  \begin{tabular}[#1]{@{}c@{}}#2\end{tabular}}

\def\ha{{\mathcal{H}}}

\def\br{{\bar{\rho}}}
\def\bp{{\bar{P}}}

\def\dr{{\delta\rho}}
\def\dP{{\delta P}}
\def\mp{{\Psi}}

\def\bk{{\mathbf{k}}}
\def\kd{{\dot{\kappa}}}
\def\icg{{\dot{\mu}}}
\def\iicg{{\mu}}
\def\dicg{{\ddot{\mu}}}

\def\DT{{\Delta^\text{T}}}
\def\DP{{\Delta^\text{P}}}
\def\Tgb{{\dot{\Theta}_{\gamma b}}}
\def\tc{{\tau_c}}
\def\nq{{\mathbf{\hat{n}}}}
\def\nn{{\nonumber}}
\def\ugc{{u_{\mathrm{DM}-\gamma}}}
\def\an{{\gamma}}

\begin{document}

\title{Constraints on $\gamma$-CDM interactions matching the Planck data precision}

\author{Julia Stadler}
\email{julia.j.stadler@durham.ac.uk}
\affiliation{Institute for Particle Physics Phenomenology, Durham University, South Road, Durham, DH1 3LE, United Kingdom}

\author{C{\'e}line B{\oe}hm}
\affiliation{Institute for Particle Physics Phenomenology, Durham University, South Road, Durham, DH1 3LE, United Kingdom}
\affiliation{LAPTH, U. de Savoie, CNRS, BP 110, 74941 Annecy-Le-Vieux, France}
\affiliation{Perimeter Institute, 31 Caroline St N., Waterloo Ontario, Canada N2L 2Y5}
\affiliation{School of Physics, University of Sydney, Camperdown, NSW 2006, Australia}

\preprint{}

\begin{abstract}
The Planck experiment has measured the Cosmic Microwave Background (CMB) angular spectrum with unprecedented accuracy, and these results are likely to remain the state-of-the art for a decade or more. Since these measurements are going to be used to forecast Euclid's ability to test alternative dark matter models, we aim to set the final CMB constraint on the cross section of dark matter-photon scattering in the early universe. As the formalism has not been published in the literature yet, we present the key steps to derive the full set of linearized Einstein and Boltzmann equations and elucidate necessary modifications to the line of sight and the tight coupling approximation. These approximation schemes are commonly used in Boltzmann codes to reduce computational costs, but the tight coupling regime has been neglected in previous works related to dark matter-photon interactions. Our computations account for a dark matter sound speed, which was neglected in previous work. We find that its effect is negligible on the CMB spectra but fairly important on the linear matter power spectrum. We then analyse the Planck data, with and without the 2015 polarisation data at high multipoles. In both cases our constraint on the ratio of the elastic scattering cross section to the dark matter mass becomes more stringent with respect to previous constraints. The change is about 35\% when the full Planck dataset is added (which includes E-mode polarisation at high multipoles and the lensing potential power spectrum) and about 20\%. when we only account for the polarisation data at low multipoles and the full temperature spectrum. Our most conservative limit reads as $\sigma_{\mathrm{DM}-\gamma} \leq 2.25\times 10^{-6}\,\sigma_\mathrm{Th}\left(m_\mathrm{DM}/\mathrm{GeV}\right)$ at 95\% confidence. We do not expect this limit to change by adding the recently published SPT data.
\end{abstract}

\maketitle

\section{Introduction}
\label{sec: intro}
The CMB angular power spectrum measurements by the Planck satellite \cite{Adam:2015rua,Ade:2013sjv} have enabled to establish the validity of the standard $\Lambda$ cold dark matter cosmological model ($\Lambda$CDM) down to galaxy scales \cite{Ade:2015xua,Ade:2015rim}. These results are now used by the Euclid \cite{Laureijs:2011gra} collaboration to forecast Euclid's ability to constrain the nature of dark energy. However, extensions of the $\Lambda$CDM scenario, allowing for interactions between dark matter and standard model particles \cite{Boehm:2000gq,Chen:2002yh,Sigurdson:2004zp,CyrRacine:2012fz,Wilkinson:2014ksa,Prinz:1998ua}, are still allowed by the data and would lead to slightly different cosmological parameters \cite{Wilkinson:2013kia}. As these alternative models predict a different linear matter power spectrum at small scales \cite{Boehm:2000gq,Boehm:2004th} than the standard $\Lambda$CDM model, they deserve to be studied with the same level of precision as $\Lambda$CDM. Indeed, Euclid and other surveys such as LSST \cite{ivezic2008lsst} and WFIRST \cite{Spergel:2015sza} may reveal deviations to the $\Lambda$CDM model predictions.

In this work we aim to constrain the presence of dark matter-photon interactions with great precision. In presence of such an interaction the linear matter power spectrum exhibits damped oscillations at small scales, which alter the evolution of the baryon-photon fluid in the early universe \cite{Boehm:2001hm}. This phenomenology can be described in terms of a single new parameter: the radio of the dark matter-photon scattering cross section to the dark matter mass, which we parametrize as
\begin{equation}
\ugc = \frac{\sigma_{\mathrm{DM}-\gamma}}{\sigma_\mathrm{Th}}~ \left(\frac{m_\mathrm{DM}}{100~\mathrm{GeV}}\right)^{-1}\,,
\label{eq: ugc}
\end{equation}
with $\sigma_\mathrm{Th}$ the Thomson cross section. In what follows, we refer to the dark matter component by using the subscript ``DM".

A constant dark matter-photon cross section, as assumed in Eq.~\ref{eq: ugc}, requires a coupling between dark matter and charged particles. Such particles could belong to the standard model or to an extension of it. There is no guarantee that dark matter couples to electrons and/or protons (for recent constraints see Ref.~\cite{Xu:2018efh, Slatyer:2018aqg}), therefore attempts to probe the dark matter-photon coupling enable one to cover all aspects of dark matter interactions.

Upper bounds on $\ugc$ have been derived from both observations of the CMB fluctuations \cite{Wilkinson:2013kia,McDermott:2010pa} and structure formation at small scales \cite{Boehm:2003xr,Boehm:2014vja,Schewtschenko:2014fca,Schewtschenko:2015rno,Escudero:2015yka,Schneider:2014rda,Dolgov:2013una}. CMB spectral distortions are also another promising avenue to derive further constraint \cite{Diacoumis:2017hff,Ali-Haimoud:2015pwa}. However, all these analyses rely on modified Boltzmann codes to either predict the CMB spectrum or obtain the primordial matter-power spectrum as initial condition for N-body simulations. 

Boltzmann codes use various approximation schemes to speed up the computations. One of them is the tight coupling approximation, which is relevant at early stages of the evolution of perturbations, well before decoupling, when baryons and photons interact significantly, i.e. both fluids are tightly coupled. The very good agreement between CMB observations and $\Lambda$CDM predictions hints that the rate for dark matter-photon scattering is much smaller than the Thomson scattering rate \cite{Boehm:2000gq,Boehm:2004th,Boehm:2001hm}. Hence, the impact of dark matter-photon scattering on the (baryonic) tight coupling approximation was thought to be negligible and has not received much attention so far. To our knowledge modified equations of the tight coupling regime, which take into account dark matter-photon scattering, have not been published in the literature. Although expressions for dipole dark matter exist \cite{Sigurdson:2004zp,Chen:2002yh}, however, these apply to dark matter-baryon scattering. Inspecting several non-public Boltzmann codes, which include dark matter-photon scattering, we found that expressions used for the tight coupling approximation varied. We therefore think it is useful to give a systematic derivation of the relevant expressions and quantify their impact on the final results.

Dark matter-photon scattering implies a non-zero dark matter sound speed. The latter was neglected in previous works on dark matter-photon scattering but its equivalent was accounted for in the context of dark matter-baryon interactions \cite{Dolgov:2013una,Chen:2002yh}. The computation of the dark matter sound speed introduces an additional parametric dependence on the dark matter mass but its inclusion in the equations is necessary to quantify the impact of the dark matter interactions on the CMB and $\rm{P(k)}$ spectra. 

Since the latest constraints of dark matter-photon interactions were derived before the CMB polarisation data in the multipole range $30\le l \le 1996$ \cite{Ade:2015xua} became available, and the equations governing the linear evolution of cosmological perturbations were never explicitly derived in the literature, we aim to clarify the formalism and provide the final constraint using the Planck data on the dark matter-photon interactions. This will become particularly relevant after the launch of the Euclid mission in 2020, designed to measure galaxy clustering and weak gravitational lensing in 36\% of the observable sky up to a redshift of $z\simeq2$ \cite{Laureijs:2011gra}. Euclid's determination of the matter power spectrum will provide a probe of $\ugc$, which is complementary to CMB constraints but extends to smaller cosmological scales and thus is potentially sensitive to even smaller values of $\ugc$.

During the preparation of this manuscript the SPTpol experiment released new measurements of the E-mode polarisation auto correlation (EE) and the temperature E-mode cross correlation (TE) angular spectrum in the multipole range $50 < l \leq 8000$ \cite{Henning:2017nuy}. These observations where obtained with a $500~\mathrm{deg}^2$ sky coverage and constitute the most sensitive measurements of the EE and TE angular spectra at large multipoles. Although these new datasets are not implemented yet in the MontePython code, which we use for parameter extraction, we have studied how they may change the limit that we obtained for the dark matter-photon scattering cross section to mass ratio.

In section \ref{sec: equations}, we recall the formalism to include dark matter-photon interactions in the equations governing the evolution of the dark mater and photon fluids. We quantify the impact that such interactions have on the cosmological parameters in section \ref{sec: parameter-extraction} and conclude in section \ref{sec: conclusions}. To derive our constraints we use a modified version of the Boltzmann code CLASS\footnote{http://class-code.net/} (version 2.6) \cite{Lesgourgues:2011re,Blas:2011rf}, which we made available online\footnote{https://github.com/bufeo/class\_v2.6\_gcdm.git}.

\section{Linear Evolution of cosmological perturbations in the presence of dark matter photon interactions}
\label{sec: equations}
In this section, we derive the full formalism, that leads to the equations, which are implemented in the CLASS code. For clarity, we first recall the equations for the $\Lambda$CDM standard model before discussing the case of dark matter-photon interactions.

Throughout the paper we will follow the notation established in \cite{Ma:1995ey}, where one finds a comprehensive derivation of the Boltzmann equations for $\Lambda$CDM. We consider a spatially flat FRW universe with comoving coordinates $x^\mu$, where $\mu$ refers to the four space-time components (roman letters running from 1 to 3 will be used to label space components). In the Newtonian gauge, scalar metric perturbations are described by two potentials $\phi$ and $\psi$ and the line element is
\begin{equation}
ds^2 = a^2(\tau) \left[-\left(1+2\psi\right)d\tau^2 + \left(1-2\phi\right)dx_i dx^i\right]\,,
\end{equation}
with $\tau$ the conformal time defined as $dx^0 = d\tau = dt/a(t)$ and $a(t)$ to the scale factor. We denote derivatives with respect to $\tau$ by a dot.

\subsection{Boltzmann hierarchy in absence of dark matter-photon interactions}
\label{subsec:bg}
We first derive the key equations that lead to the well-known perturbation equations associated with the relativistic and non-relativistic species in a $\Lambda$CDM scenario. We start by decomposing the stress energy tensor $T^\mu_\nu$ into a homogeneous background and small fluctuations around it, i.e. $T^\mu_\nu = \bar{T}^\mu_\nu + \delta T^\mu_\nu$. From this, it is straightforward to derive that the density perturbation $\delta$, the velocity dispersion $\theta$ and the shear stress $\sigma$ of these species are related to the stress energy tensor via the following expressions
\begin{subequations}
\begin{align}
&\delta \equiv \delta\rho/\bar{\rho} = \delta T^0_0/\bar{T}^0_0\,,\\
&\left(\bar{\rho}+\bar{P}\right)\theta \equiv i\, k^i\,\delta T^0_i\,,\\
&\left(\bar{\rho}+\bar{P}\right)\sigma \equiv -\left(\hat{k}_i\hat{k}^j - \frac{1}{3}\delta_i^j \right)\left(T^i_j - \frac{1}{3}\delta^i_j T^k_k\right)\,,
\end{align}
\end{subequations}
where $P$ and $\rho$ denote the pressure and energy density respectively. The bar symbol refers to background quantities while the hat indicates unit vectors.

\subsubsection{Perturbations for non-relativistic species}
The evolution of the baryon and dark matter perturbations follows from the conservation of the stress-energy tensor $\nabla_\mu T^\mu_\nu = 0$, where $\nabla_\mu$ is the covariant derivative of the perturbed metric. Expanding the conservation equation to first order in perturbations yields
\begin{subequations}
\begin{align}
\dot{\delta} &= -3\ha\left(c_s^2 - w\right) - (1+w)\left(\theta-3\dot{\phi}\right)\,,
\label{eq: em-conservation-single-delta}\\
\dot{\theta} &= -\ha(1-3w)\theta - \frac{\dot{w}}{1+w}\theta + \frac{\dP/\dr}{1+w}\,k^2\delta\nn\\
&\quad- k^2\sigma + k^2\psi\,,
\label{eq: em-conservation-single-theta}
\end{align}
\label{eq: em-conservation-single}
\end{subequations}
where we defined the adiabatic sound speed as $c_s^2 \equiv dP/d\rho \equiv w + \rho w/d\rho$ and the reduced Hubble rate $\ha\equiv\dot{a}/a$. Eq.~(\ref{eq: em-conservation-single}) is valid for a single uncoupled fluid but needs to be modified for a species, which interacts with another component. We will derive these modifications for baryons coupled to photons below but first focus on dark matter.

In the $\Lambda$CDM scenario dark matter is described by a perfect fluid with vanishing sound speed and shear stress. The possible dark matter interactions with Standard Model particles (or within an hypothetical dark sector) are disregarded. Under this assumption, Eq.~(\ref{eq: em-conservation-single}) reduces to
\begin{subequations}
\begin{align}
\dot{\delta}_\mathrm{DM} &= -\theta_\mathrm{DM} + 3\dot{\phi}\,,\\
\dot{\theta}_\mathrm{DM} &= -\ha\theta_\mathrm{DM} + k^2\psi\,.
\end{align}
\end{subequations}

For baryons energy-momentum conservation needs to be considered at the level of the coupled baryon-photon fluid. Thus Eq.~(\ref{eq: em-conservation-single}) is valid for the mass-averaged perturbations $\delta$ and $\theta$ of the coupled fluids, i.e. Eq.~(\ref{eq: em-conservation-single-theta}) becomes
\begin{align}
&\sum_s \left( \br_s + \bp_s \right)\dot{\theta}_s =\nn\\
&\quad-\ha\sum_s \br_a(1+w_s)(1-3w_s)\theta_s - \sum_s \dot{w}_s\br_s\theta_s\nn\\
&\quad+ k^2\sum_s \frac{\dP_s}{\dr_s}\br_s\delta_s- k^2\sum_s\br_s(1+w_s)\sigma_s \nn\\
&\quad+ k^2\psi\sum_s\br_s(1+w_s)\,,
\label{eq: em-conservation-multi-theta}
\end{align}
where in the sum $s$ runs over baryons and photons. Using Eq.~(\ref{eq: boltzmann-h-photons}) below for the photon perturbations the evolution of baryons is then described by
\begin{subequations}
\begin{align}
\dot{\delta}_b &= -\theta_b + 3\dot{\phi}\,,\\
\dot{\theta}_b &= -\ha\theta_b + k^2 c_b^2\delta_b + k^2\psi - R \kd\left(\theta_b - \theta_\gamma\right)\,,
\label{eq: boltzmann-theta_b_dot}
\end{align}
\end{subequations}
where $R\equiv4\rho_\gamma/3\rho_b$ and $c_b^2$ is the baryon sound speed.

\subsubsection{Perturbations for relativistic species}
Now, that we have derived the Boltzmann equations for all non-relativistic components, we can turn to the evolution of photons and neutrinos. For these relativistic species the shear stress can not be neglected and a description in terms of the full phase space distribution is needed. The phase space distribution can be expanded in terms of a small perturbation $\mp$ around the homogeneous background distribution $f_0$
\begin{equation}
f(x^i,q,n_j,\tau) = f_0(q)\left[1+ \mp(x^i,q,n_j,\tau)\right]\,.
\end{equation}
Depending on the species that is considered $f_0$ either follows a Bose-Einstein distribution (photons) or a Fermi-Dirac statistic (neutrinos) with temperatures $T_\gamma$ or $T_\nu$ respectively. The energy-momentum tensor is related to the phase space distribution by the integral
\begin{equation}
T_{\mu\nu} = \int \left(\prod_{k=1}^{3} dP_k\right) \left(-g\right)^{-1/2}\frac{P_\mu P_\nu}{P_0}\, f(x^i, q, n_j, \tau)\,,
\end{equation}
where $g$ is the determinant of the (perturbed) metric tensor with $(-g)^{-1/2} = a^{-4}(1-\psi+3\phi)$. $\mathbf{q}$ is the comoving proper momentum, which we decompose into modulus $q=|\mathbf{q}|$ and direction $n_i = q_i/q$. The evolution of $f(x^i,q,n_j,\tau)$ is then governed by the Boltzmann equation
\begin{equation}
\frac{Df}{d\tau} = \frac{\partial f}{\partial\tau} + \frac{dx^i}{d\tau}\frac{\partial f}{\partial x^i} + \frac{dq}{d\tau}\frac{\partial f}{\partial q} + \frac{dn_i}{d\tau}\frac{\partial f}{\partial n_i} = \left(\frac{\partial f}{\partial\tau}\right)_C\,,
\label{eq: boltzmann}
\end{equation}
where the right hand side represents a collision term, accounting for interactions between individual species. Expanding Eq.~(\ref{eq: boltzmann}) in perturbations, collecting all terms, which are first order and finally applying a Fourier transformation yields a differential equation for the evolution of $\mp$
\begin{equation}
\frac{\partial\mp}{\partial\tau} + \frac{q}{\epsilon}i\left( \mathbf{\hat{n}} \cdot \mathbf{k}\right)\mp + \frac{\partial \ln f_0}{\partial \ln q}\left( \dot{\phi} - i\frac{\epsilon}{q}\mathbf{\hat{n}}\cdot\mathbf{k}~\psi\right) = \frac{1}{f_0}\left(\frac{\partial f_0}{\partial\tau}\right)_C\,,
\label{eq: boltzmann-intermdiate-fpert}
\end{equation}
where $\epsilon = (g^2 + a^2m^2)^{1/2}$ is the proper energy measured by a comoving observer.

We can now focus on the particular case of the photons and massless neutrinos, which provides a good description of the evolution of neutrino perturbations during the decoupling. We first get rid off the momentum dependence of the energy distribution $f(x^i,q,n_j,\tau)$ by making a Legendre polynomial decomposition of the quantity
\begin{align}
F_s(\bk, \nq,\tau) &\equiv \frac{\int q^3 dq\, f_{s,0}(q)\,\mp_s}{\int q^3 dq\, f_{s,0}(q)}\,,
\end{align}
(with $s=\gamma,\nu$). Using the expression of $F_s(\bk, \nq,\tau)$ thus obtained, namely 
\begin{align}
F_s(\bk, \nq,\tau) &\equiv \sum_{l=0}^\infty (-i)^l(2l+1)\, F_{s\, l}(\bk,\tau)\,P_l\,,
\label{eq: Fn-Legendre-expansion}
\end{align}
we then express the density fluctuation, velocity dispersion and shear stress as 
\begin{equation}
\delta = F_0\,,\quad \theta = \frac{3k}{4}F_1\,, \quad \sigma = \frac{1}{2}F_2\,. 
\end{equation}

The evolution of the cosmological perturbations is particularly simple in the case of neutrinos since the collision term can be neglected, owing to the very weak strength of the neutrino interactions at very low energy. Integrating Eq.~(\ref{eq: boltzmann-intermdiate-fpert}) over $\int dq \, q^3 f_0(q)$ and dividing by the same integral gives the evolution of $F_\nu$
\begin{equation}
\dot{F_\nu} +ik\mu F_\nu = 4\left( \dot{\phi} -ik\mu\psi\right)\,.
\label{eq: boltzmann-neutrinos-fullF}
\end{equation}
To get the evolution of the energy density, velocity dispersion, and shear we then need to substitute $F_\nu$ by its Legendre expansion (as expressed in Eq.~(\ref{eq: Fn-Legendre-expansion})), integrate over $\int_{-1}^1 dx P_n(x)$ where $n=0,1,2...$ and make use of the orthogonality relations. This finally yields the following set of coupled differential equations where the evolution of $\dot{F}_{\nu l}$ is related to the evolution of the $l+1$ and $l-1$ multipoles only:
\begin{subequations}
\begin{align}
\dot{\delta}_\nu &= 4\dot{\phi} - \frac{4}{3}\theta_\nu\,,\\
\dot{\theta}_\nu &= k^2\left(\frac{\delta_\nu}{4} -\sigma_\nu\right) + k^2\psi\,,\\
\dot{F}_{\nu l}  &= \frac{k}{2l+1} \left[l F_{\nu (l-1)}  -(l+1) F_{\nu (m+1)}\right]\,,\quad l\ge2\,.
\end{align}
\end{subequations}

The evolution of the photon fluid is different because the collision term (namely the Thomson interactions) can no longer be neglected. Owing to the expression of the cross section these interactions introduce a polarisation dependence, which can be accounted for by simultaneously evolving both the sum ($F_\gamma(\bk,\nq,\tau) $) and the difference ($G_\gamma(\bk,\nq,\tau)$) of the phase space densities associated with each photon polarisation states. Denoting $\mp_1$ and $\mp_2$ the perturbations of each polarisation state, we can define
\begin{subequations}
\begin{align}
F_\gamma &= \frac{\int dq d^3\, f_0(q) \left( \mp_1 + \mp_2 \right)}{\int dq d^3\, f_0(q)}\,,\\
G_\gamma &= \frac{\int dq d^3\, f_0(q) \left( \mp_1 - \mp_2 \right)}{\int dq d^3\, f_0(q)}\,,
\end{align}
\end{subequations}
which leads to the corresponding Boltzmann equations 
\begin{subequations}
\begin{align}
\dot{F}_\gamma + ik\mu F_\gamma - 4\left(\dot{\phi} - ik\mu\psi\right) &= \left(\frac{\partial F_\gamma}{\partial \tau}\right)_C\,,
\label{eq: boltzmann-photons-fullF}\\
\dot{G}_\gamma + ik\mu G_\gamma  &= \left(\frac{\partial G_\gamma}{\partial \tau}\right)_C\,.
\label{eq: boltzmann-photons-fullG}
\end{align}
\label{eq: boltzmann-photons-full}
\end{subequations}
The respective collision terms are given by \cite{Bond:1984fp,Kosowsky:1994cy}
\begin{subequations}
\begin{align}
&\left( \frac{\partial F_\gamma}{\partial \tau} \right)_{C} = \kd \left[ -F_\gamma + F_{\gamma 0} + 4\hat{n}\cdot \mathbf{v_e}
\right.\nn\\
&\quad\left.
- \frac{1}{2} \left( F_{\gamma 2} + G_{\gamma 0} + G_{\gamma 2} \right) P_2 \right]\,,
\label{eq: collision-gb-F}\\
&\left(\frac{\partial G_\gamma}{\partial \tau}\right)_C = \kd \left[ -G_\gamma + \frac{1}{2}\left( F_{\gamma 2} + G_{\gamma 0} + G_{\gamma 2} \right) \left(1-P_2\right)\right]\,,
\end{align}
\end{subequations}
where $\mathbf{v}_e$ is the electron's coordinate three velocity related to the velocity dispersion by $i k_j v^j = \theta$\,, and the Thomson scattering rate is defined as $\kd=a n_e\sigma_\mathrm{Th}$. Integrating Eq.~(\ref{eq: boltzmann-photons-full}) against Legendre polynomials in the same fashion as done for neutrinos we obtain a hierarchy of coupled equations
\begin{subequations}
\begin{align}
\dot{\delta}_\gamma &= -\frac{4}{3}\theta_\gamma + 4\dot{\phi}\,,
\label{eq: boltzmann-h-F0}\\
\dot{\theta}_\gamma &= k^2\left(\frac{1}{4}\delta_\gamma - \sigma_\gamma\right) + k^2\psi + \kd\left(\theta_b - \theta_\gamma\right)
\label{eq: boltzmann-h-F1}\,,\\
\dot{\sigma}_\gamma &= \frac{4}{15}\theta_\gamma - \frac{3}{10}k F_{\gamma 3} - \frac{9}{10}\kd\sigma_\gamma + \frac{1}{20}\kd\left(G_{\gamma 0} + G_{\gamma 2}\right)\,,
\label{eq: boltzmann-h-F2}\\
\dot{F}_{\gamma l} &= \frac{k}{2l+1}\left[l F_{\gamma (l-1)} - (l+1)F_{\gamma(l+1)}\right] - \kd F_{\gamma l}\,, \quad l\ge3\,,
\label{eq: boltzmann-h-Fge3}\\
\dot{G}_{\gamma 0} &= -kG_{\gamma 1} - \frac{1}{2}\kd\left(G_{\gamma 0} - F_{\gamma 2} - G_{\gamma 2}\right)\,,
\label{eq: boltzmann-h-G0}\\
\dot{G}_{\gamma 1} &= \frac{k}{3}\left( G_{\gamma 0} - 2G_{\gamma 2}\right) - \kd G_{\gamma 1}\,,
\label{eq: boltzmann-h-G1}\\\
\dot{G}_{\gamma 2} &= \frac{k}{5}\left( 2G_{\gamma 1} - 3 G_{\gamma 3} \right) + \frac{\kd}{10} \left(F_{\gamma 2} + G_{\gamma 0} + G_{\gamma 2}\right)\nn\\
&- \kd G_{\gamma 2}\,,
\label{eq: boltzmann-h-G2}\\\
\dot{G}_{\gamma l} &= \frac{k}{2l + 1} \left( l G_{\gamma (l-1)} - (l+1) G_{\gamma(l+1)}\right) - \kd G_{\gamma l}\,,\quad l\ge3\,.
\label{eq: boltzmann-h-Gge3}
\end{align}
\label{eq: boltzmann-h-photons}
\end{subequations}

\subsection{Boltzmann Hierarchy in presence of dark matter-photon interactions}
The introduction of dark-matter-photon interactions changes both the dark matter and photon Boltzmann equations. This was explained in details in \cite{Boehm:2001hm,Wilkinson:2013kia}. Here we recall the main equations for the photon and the dark matter components.

A collisional term, that reflects the interactions between the Dark Matter and the photons, needs to be added to the right hand side of Eq.~(\ref{eq: boltzmann-photons-fullF}) and Eq.~(\ref{eq: boltzmann-photons-fullG}). Here we model it in the same way as for the baryon-photon interactions. I.e. we assume that the dark matter-photon scattering amplitude has the same angular polarisation dependence as the Thomson elastic scattering cross section. We leave the investigation of different angular dependencies for further studies. Under these assumptions we find
\begin{subequations}
\begin{align}
\left( \frac{\partial F_\gamma}{\partial \tau} \right)_{\mathrm{DM}-\gamma} &\hspace{-10pt}= \icg \left[ -F_\gamma + F_{\gamma 0} + 4\hat{n}\cdot \mathbf{v_\mathrm{DM}}
\right.\nn\\
&\qquad\left.- 
\frac{1}{2} \left( F_{\gamma 2} + G_{\gamma 0} + G_{\gamma 2} \right) P_2\right]
\label{eq: collision-gdm-F}\,,\\
\left(\frac{\partial G_\gamma}{\partial \tau}\right)_{\mathrm{DM}-\gamma} &\hspace{-10pt}= \icg \left[ -G_\gamma + \frac{1}{2}\left( F_{\gamma 2} + G_{\gamma 0} + G_{\gamma 2} \right) \left(1-P_2\right)\right]\,,
\label{eq: collision-gdm-G}
\end{align}
\end{subequations}
with $\icg$ being the rate for dark matter-photon scattering $\icg\equiv a n_\mathrm{DM} \sigma_{\mathrm{DM}-\gamma}$. With this addition the Boltzmann equations for $F_\gamma$ and $G_\gamma$ become
\begin{subequations}
\begin{align}
&\dot{F}_\gamma + ik\mu F_\gamma = 4\left(\dot{\phi} - ik\mu\psi\right)\nn\\ 
&\quad+ \left(\kd+\icg\right)\left[-F_\gamma + F_{\gamma 0} - \frac{1}{2}\left(F_{\gamma 2} + G_{\gamma 0} + G_{\gamma 2}\right)P_2(\mu)\right]\nn\\
&\quad+4\kd\,\nq\cdot\mathbf{v_e} +4\icg\,\nq\cdot\mathbf{v_\mathbf{DM}}\,,
\label{eq: boltzmann-eq-F-icg}\\
&\dot{G}_\gamma + ik\mu G_\gamma\nn\\
&= \left(\kd + \icg\right)\left[ -G_\gamma + \frac{1}{2}\left( F_{\gamma 2} + G_{\gamma 0} + G_{\gamma 2} \right) \left(1-P_2(\mu)\right)\right]\,.
\label{eq: boltzmann-eq-G-icg}
\end{align}
\label{eq: boltzmann-full-gcdm-phtons}
\end{subequations}%
The decomposition into equations for the individual multipoles is similar to that in Sec.~\ref{subsec:bg} and also yields a set of coupled equations. However, they are augmented by a new interaction term:
\begin{subequations}
\begin{align}
\dot{\delta}_\gamma &= -\frac{4}{3}\theta_\gamma + 4\dot{\phi}\,,
\label{eq: boltzmann-h-cg-F0}\\
\dot{\theta}_\gamma &= k^2\left(\frac{1}{4}\delta_\gamma-\sigma_\gamma\right) + k^2\psi + \kd\left(\theta_b-\theta_\gamma\right) + \icg\left(\theta_\mathrm{DM}-\theta_\gamma\right)\,,
\label{eq: boltzmann-h-cg-F1}\\
\dot{\sigma}_\gamma &= \frac{4}{15}\theta_\gamma - \frac{3}{10}k F_{\gamma 3} - \frac{9}{10}\left(\kd+\icg\right)\sigma_\gamma\nn\\
&\quad+ \frac{1}{20}\left(\kd+\icg\right)\left(G_{\gamma 0} + G_{\gamma 2}\right)\,,
\label{eq: boltzmann-h-cg-F2}\\
\dot{F}_{\gamma l} &= \frac{k}{2l+1}\left[l F_{\gamma (l-1)} - (l+1)F_{\gamma(l+1)}\right]\nn\\
&\quad- \left(\kd+\icg\right) F_{\gamma l}\,, \quad l\ge3\,,
\label{eq: boltzmann-h-cg-Fge3}\\
\dot{G}_{\gamma 0} &= -k G_{\gamma 1} - \frac{1}{2}\left(\kd+\icg\right)\left(G_{\gamma 0} - F_{\gamma 2} - G_{\gamma 2}\right)\,,
\label{eq: boltzmann-h-cg-G0}\\
\dot{G}_{\gamma 1} &= \frac{k}{3}\left( G_{\gamma 0} - 2G_{\gamma 2}\right) - \left(\kd+\icg\right) G_{\gamma 1}\,,
\label{eq: boltzmann-h-cg-G1}\\
\dot{G}_{\gamma 2} &= \frac{k}{5}\left( 2G_{\gamma 1} - 3 G_{\gamma 3} \right) + \frac{\kd+\icg}{10} \left(F_{\gamma 2} + G_{\gamma 0} + G_{\gamma 2}\right)\nn\\
&- \left(\kd+\icg\right) G_{\gamma 2}\,,
\label{eq: boltzmann-h-cg-G2}\\
\dot{G}_{\gamma l} &= \frac{k}{2l + 1} \left( l G_{\gamma (l-1)} - (l+1) G_{\gamma(l+1)}\right)\nn\\
&\quad- \left(\kd+\icg\right) G_{\gamma l}\,,\quad l\ge3\,.
\label{eq: boltzmann-h-cg-Gge3}
\end{align}
\label{eq: boltzmann-h-gcdm-photons}
\end{subequations}

\subsection{Dark matter evolution in the presence of dark matter-photon interactions}

While the evolution equations of baryons and neutrinos remain unaffected by dark matter-photon scattering, the evolution of dark matter perturbations is modified according to Eq.~(\ref{eq: em-conservation-multi-theta}), yielding
\begin{subequations}
\begin{align}
\dot{\delta}_\mathrm{DM} &= -\theta_\mathrm{DM} + 3\dot{\phi}\,,
\label{eq: delta-cdm-icg}\\
\dot{\theta}_\mathrm{DM} &= k^2\psi - \ha\theta_\mathrm{DM} + c_\mathrm{DM}k^2\delta_\mathrm{DM} -S\icg\left(\theta_\mathrm{DM} - \theta_\gamma\right)\,,
\label{eq: theta-dm-icg}
\end{align}
\label{eq: boltzmann-h-gcdm-cdm}
\end{subequations}
where $S\equiv4\rho_\gamma/3\rho_\mathrm{DM}$ and we have allowed for a non-zero dark matter sound speed \cite{Ma:1995ey,Chen:2002yh}
\begin{equation}
c_\mathrm{DM}^2 = \frac{k_\mathrm{B}\,T_\mathrm{DM}}{m_\mathrm{DM}}\left[1 - \frac{1}{3}\,\frac{\partial\ln T_\mathrm{DM}}{\partial\ln a} \right]\,, \label{sound_speed} 
\end{equation}
with $m_\mathrm{DM}$ being the mass of the dark matter particle.\\

In the standard $\Lambda$CDM scenario dark matter is assumed to decouple from the photon bath at some early time after which $T_\mathrm{DM}$ evolves as $a^{-2}$. Consequently the dark matter temperature is small at all times relevant to the computation of CMB fluctuations and hence the dark matter sound speed can be neglected. Elastic scattering between dark matter and photons, however, maintains kinetic equilibrium between both species and implies a larger dark matter temperature, which now evolves as
\begin{equation}
\dot{T}_\mathrm{DM} = -2\ha\,T_\mathrm{DM} + \frac{8\,\rho_\gamma}{3\,\rho_\mathrm{DM}}\icg\left(T_\gamma - T_\mathrm{DM} \right)\,. \label{Tdot_dm}
\end{equation}

In light of this the effect of the dark matter sound speed on the computed CMB spectra can not be neglected a priori, as done in previous works on dark matter-photon scattering \cite{Wilkinson:2013kia}. However, as we show in section \ref{sec: impact}, for realistic dark matter masses the inclusion of the dark matter sound speed affects the computation of CMB spectra at a negligible level of accuracy.

\subsection{Line of Sight Integration}
\label{sesec: eq-los}
Now that we have derived the new set of equations, we can focus on their resolution and the treatment of the baryon-photon tight coupling regime. The computation time, required to solve the Boltzmann equations, can be significantly reduced \cite{Bond:1984fp} by introducing the line of sight integration approach \cite{Seljak:1996is,Hu:1997hp}. Here we derive all modifications to this method necessitated by the presence of dark matter-photon interactions. Starting point for this derivation is Eq.~(\ref{eq: boltzmann-full-gcdm-phtons}), which, defining ${\Gamma \equiv F_{\gamma 2} + G_{\gamma 0} + G_{\gamma 2}}$, $\DT\equiv F_\gamma/4$, $\DP\equiv G_\gamma/4$, $\nq\cdot\mathbf{v_e} \equiv\an v_b$, and $\nq\cdot\mathbf{v_\mathrm{DM}}\equiv \an v_\mathrm{DM}$, becomes
\begin{align}
&\dot{\DT} + ik\an \DT = \left(\dot{\phi} - ik\an\psi\right)\nn\\
&\quad+ \left(\kd+\icg\right)\left[ -\DT + \DT_0 - \frac{1}{8}\Gamma P_2 \right] +\kd\an v_b + \icg\an v_\mathrm{CM}
\label{eq: los-initial-T}\,,\\
&\dot{\DP} + ik\an\DP = \left(\kd+\icg\right)\left[-\DP + \frac{1}{8}\left(1-P_2\right)\Gamma \right]\,.
\label{eq: los-initial-P}
\end{align}
We now define
\begin{equation}
\kappa \equiv -\int_{\tau_0}^\tau d\tau\, \kd(\tau)\,, \quad \text{and} \quad \iicg \equiv -\int_{\tau_0}^\tau d\tau\, \icg(\tau)\,,
\label{eq: los-def-kappa-and-mu}
\end{equation}
and the visibility function $g(\tau)$
\begin{equation}
g(\tau) \equiv \left(\kd+\icg\right)e^{-\kappa-\iicg}\,,
\label{eq: los-def-visibility-function-icg}
\end{equation}
which describes the probability that a photon last scattered between $\tau$ and $d\tau$ and thus is sharply peaked at decoupling. 

We can simplify the polarisation equation by integrating $\partial_t\left(\DP \exp(ik\an-\kappa-\iicg) \right)$ over the past lightcone from $0$ to the present time $\tau_0$, which leads to
\begin{equation}
\DP =\frac{3}{16}\int_0^{\tau_0} d\tau~ g(\tau) e^{ik\an(\tau-\tau_0)}\left(1-\an^2\right)\Gamma\,.
\label{eq: los-3-P}
\end{equation}
The previous equation was obtained by integration by parts where the boundary terms vanish. Finally, the dependence on the angle $\gamma$ can be removed by again integrating by parts, leading to 
\begin{equation}
\DP =\frac{3}{16}\int_0^{\tau_0}d\tau\,e^{ik\an(\tau-\tau_0)}\left[ g\Gamma + \frac{1}{k^2}\frac{d^2}{d\tau^2}\left(g\Gamma\right) \right]\,.
\label{eq: los-4-P}
\end{equation}
Using the same tricks enables us to express $\DT$ as 
\begin{align}
\DT(\tau_0) &=\int_0^{\tau_0} d\tau\, e^{ik\mu(\tau-\tau_0)} \left\{ e^{-\kappa-\icg}\dot{\phi} + \frac{g}{4}\left( \delta_\gamma + \frac{1}{4}\Gamma\right)\right.\nn\\
&+ \frac{d}{d\tau}\left[ e^{-\kappa-\iicg}\psi \right] -\frac{e^{-\kappa-\iicg}}{ik}\left( \ddot{\kappa}\,v_b + \kd\left(\kd+\icg\right)\,v_b \right. \nn\\
&\left.+ \kd\,\dot{v}_b + \dicg\, v_\mathrm{DM} + \icg(\kd+\icg)\,v_\mathrm{DM} + \icg\, \dot{v}_\mathrm{DM} \right) \nn\\
&\left.+ \frac{3}{16k^2}\frac{d}{d\tau}\left(\dot{g}\Gamma + g\dot{\Gamma} \right) \right\}\,.
\label{eq: los-1-T}
\end{align}

We can now define the source functions $S^T$ and $S^P$ as
\begin{equation}
\Delta^\text{T,P} (k,\tau_0) = \int_0^{\tau_0} d\tau\, e^{ik\mu(\tau-\tau_0)} S^\text{T,P}(k,\tau)\,.
\label{eq: los-source-definition}
\end{equation}
The final step of the line of sight integration consists in expanding the exponential in Eq.~(\ref{eq: los-source-definition}) in terms of its radial and angular eigenfunctions, which are spherical Bessel functions and Legendre polynomials respectively. Performing the ensemble average and integrating over $\an$ allows us to express the multipoles $\DT_l$ and $\DP_l$ as an integral over a source and a spherical Bessel function 
\begin{equation}
\Delta_l^\text{T,P}(k) = \int_0^{\tau_0}d\tau\, S^\text{T,P}(k,\tau)\, j_l\left[k\left(\tau_0-\tau\right) \right]\,.
\end{equation}
The comparison of Eq.~(\ref{eq: los-source-definition}) with Eq.~(\ref{eq: los-4-P}) and Eq.~(\ref{eq: los-1-T}) yields an expression for the source functions, which only depends on the first few multipoles and thus simplifies the computational effort tremendously
\begin{subequations}
\begin{align}
&S^\text{T}(k,\tau) = e^{-\kappa-\iicg}\dot{\phi} + \frac{g}{4}\left(\delta_\gamma + \frac{1}{4}\Gamma\right)+ \frac{e^{-\kappa-\iicg}}{k^2}\times\nn\\
&\quad\times\left\{ \left[\ddot{\kappa} + \kd\left(\kd+\icg\right) \right]\theta_b + \kd\,\dot{\theta}_b + \left[\dicg + \icg\left(\icg+\kd\right)\right]\theta_\mathrm{DM} \right.\nn\\
&\quad\left.+\,\icg\,\dot{\theta}_\mathrm{DM}\right\}+\frac{d}{d\tau}\left[ e^{-\kappa-\iicg}\psi + \frac{3}{16k^2}\left(\dot{g}\Gamma + g\dot{\Gamma} \right) \right]\,,
\label{eq: los-sourceT-icg}\\
&S^\text{P}(k,\tau) = \frac{3}{16}\left[ g\Gamma + \frac{1}{k^2}\frac{d^2}{d\tau^2}\left(g\Gamma\right)\right]\,.
\label{eq: los-sourceP-icg}
\end{align}
\end{subequations}
We note that $S^P$ is modified by the dark matter-photon interactions only indirectly, through the definition of the visibility function $g$, while the dark matter-photon scattering rate enters the expressions for $S^T$ directly.

In CLASS the temperature source function is split into three different parts \cite{Lesgourgues:2013bra}
\begin{equation}
S_T = {S}_T^0 + \frac{d}{d\tau}{S}_T^1 + \frac{d}{d\tau}\dot{{S}}_T^2\,,
\label{eq: los-splitting-old}
\end{equation}
where the dot indicates that the code evaluates the analytic expression for the derivative. The derivatives that are denoted by $d/d\tau$ are estimated numerically. In the past this splitting has led to some confusion in the context of dark matter-photon interactions. It is therefore useful to provide the explicit expressions for all three terms
\begin{subequations}
\begin{align}
S_0^T &= g\left(\frac{1}{4}\delta_\gamma+\phi\right) + 2 e^{-\kappa-\iicg}\dot{\phi} + \frac{1}{k^2}\left[ g\left(\kd\theta_b+\icg\theta_\mathrm{DM} \right)\right.\nn\\
&\left.\quad+ e^{-\kappa-\icg}\left(\ddot{\kappa}\,\theta_b + \dicg\,\theta_\mathrm{DM} \right)
+ e^{-\kappa-\iicg}\left(\kd\,\dot{\theta}_b + \icg\,\dot{\theta}_\mathrm{DM} \right) \right]\,,\rule[-1.em]{0pt}{0pt}
\label{eq: code-los-ST0}\\
S_T^1 &= e^{-\kappa-\iicg}\,k\left(\psi-\phi \right)\,,
\label{eq: code-los-ST1}\\
S_T^2 &= \frac{1}{8}\,g\,\Gamma\,.
\label{eq: code-los-ST2}
\end{align}
\end{subequations}

\subsection{Tight Coupling Approximation}
\label{sesec: eq-tca}
At early times the interaction rates in Eq.~(\ref{eq: boltzmann-h-gcdm-photons}) and Eq.~(\ref{eq: boltzmann-theta_b_dot}) become large and hamper numerical integration. In this regime photons and baryons are tightly coupled such that $\theta_\gamma - \theta_b$ is very close to zero. It is possible to reformulate the evolution equations in terms of the photon baryon slip
\begin{equation}
\Theta_{\gamma b} \equiv \theta_\gamma - \theta_b\,,
\label{eq: tca-def-slip}
\end{equation}
for which an analytic expansion, in powers of the inverse Thomson scattering $\tc\equiv\kd^{-1}$ rate, can be found \cite{Peebles:1970ag,Ma:1995ey}. This approach eliminates all occurrences of $\kd$ from the evolution equations and renders numerical integration feasible. 

Experimental constraints indicate that the dark matter-photon scattering rate $\icg$ has to be much smaller than $\kd$. This means that the expansion that we performed stays valid. We will comment on this in more detail below. However, the main point is that the equations governing the photon-baryon slip and the evolution of cosmological perturbations are modified in the presence of dark matter-photon interactions. Here we derive all necessary changes following the steps of Ref.~\cite{Blas:2011rf}.

To obtain equations, which are easier to integrate numerically, $\kd$ can be removed from the evolution of $\theta_b$ and $\theta_\gamma$ by rewriting them as
\begin{align}
&\dot{\theta}_b = -\frac{1}{1+R}\left[\ha\theta_b - c_b^2k^2\delta_b - k^2R\left(\frac{1}{4}\delta_\gamma - \sigma_\gamma \right) \right.\nn\\
&\quad\left.+ R\Tgb+ \icg R\left(\theta_\gamma-\theta_\mathrm{DM} \right) \right] + k^2\psi\,,
\label{eq: tca-icg-theta_b}\\
&\dot{\theta}_\gamma = -R^{-1}\left[\ha\theta_b - c_b^2k^2\delta_b + \dot{\theta}_b \right] + k^2\left(\frac{1}{4}\delta_\gamma - \sigma_\gamma \right)\nn\\
&\quad+ \frac{1+R}{R}k^2\psi - \icg\left(\theta_\gamma - \theta_\mathrm{DM}\right)\,.
\label{eq: tca-icg-theta_g}
\end{align}
At the same time, Eq.~(\ref{eq: boltzmann-theta_b_dot}) and Eq.~(\ref{eq: boltzmann-h-cg-F1}) can be combined to give an implicit equation for $\Theta_{\gamma b}$
\begin{align}
&\tc\left[\Tgb - \ha\theta_b + k^2\left(c_b^2\delta_b - \frac{1}{4}\delta_\gamma + \sigma_\gamma \right) + \icg\left(\theta_\gamma - \theta_\mathrm{DM}\right) \right]\rule[-1.em]{0pt}{0pt}\nn\\
&\quad+(1+R)\Theta_{\gamma b} = 0\,.
\label{eq: tca-slip-starting-point}
\end{align}
This equation can be cast into the form
\begin{equation}
\epsilon \dot{y}(t) + y(t)/f(t) + \epsilon g(t) = 0\,,
\label{eq: tca-y-equation}
\end{equation}
where $\epsilon\ll1$. Using a sum series ansatz $y(t) = \sum_{n=1} \epsilon^n y_n(t)$, the solution up to second order in $\epsilon$ is found to be
\begin{equation}
y = -\epsilon f g + \epsilon^2 f\left( \dot{f}g + f\dot{g} \right)\,.
\label{eq: tca-schematic-solution-o2}
\end{equation}
To make the analogy we multiply Eq.~(\ref{eq: tca-slip-starting-point}) by $\bar{\tc}/\tc$, where $\bar{\tc}$ is the opacity at some arbitrary time, around which we perform the expansion. We require that the dimensionless parameters $\tc\ha$, $\tc k$ and $\tc\icg$ are all much smaller than unity. Now $\dot{\bar{\tc}} = 0$ and we can identify
\begin{align}
{f}_\Theta &= \frac{\tc}{\bar{\tc}}\frac{1}{1+R}\,,\\
g_\Theta &= -\ha\theta_b + k^2\left( c_b^2\delta_b - \frac{1}{4}\delta_\gamma + \sigma_\gamma \right) + \icg\left(\theta_\gamma-\theta_\mathrm{DM}\right)\,.
\label{eq: tca-icg-gTheta}
\end{align}
Using the result of Eq.~(\ref{eq: tca-schematic-solution-o2}), we find an expression for $\Theta_{\gamma b}$ up to second order in $\bar{\tc}$
\begin{align}
\Theta_{\gamma b} &= -\bar{\tc}\bar{f}_\Theta g + \bar{\tc}^2\bar{f}_\Theta\left(\dot{\bar{f}}_\Theta g_\Theta + \bar{f}_\Theta\dot{g}_\Theta\right) + \mathcal{O}(\tc^3)\nn\\
&= -\tilde{f}_\Theta g_\Theta + \tilde{f}_\Theta\left( \dot{\tilde{f}}_\Theta g_\Theta + \tilde{f}_\Theta \dot{g}_\Theta  \right)+ \mathcal{O}(\tc^3)\,,
\end{align}
where, in the second line, we defied $\tilde{f}_\Theta = \tc/(1+R)$. Eventually we are interested in an expression for the photon-baryon slip $\Tgb$, which we now write as
\begin{align}
\Tgb 
&= \left(\frac{\dot{\tilde{f}}_\Theta}{\tilde{f}_\Theta} \right) \Theta_{\gamma b} + \tilde{f}_\Theta\left(-\dot{g}_\Theta + \ddot{\tilde{f}}_\Theta g_\Theta + 2\dot{\tilde{f}}_\Theta \dot{g}_\Theta + \tilde{f}_\Theta \ddot{g}_\Theta \right)\nn\\
&\quad+ \mathcal{O}(\tc^3)\,.
\label{eq: tca-slip-schematic-o2}
\end{align}
Eq.~(\ref{eq: tca-slip-schematic-o2}) requires the computation of $\dot{g}_\Theta$, which is conventionally cast in the form
\begin{align}
&\dot{g}_\Theta  
=2\ha\Tgb - \frac{\ddot{a}}{a}\theta_b + k^2\left[ \bar{c_b^2}\delta_b + c_b^2\dot{\delta}_b - \frac{1}{4}\dot{\delta}_\gamma + \dot{\sigma}_\gamma - \ha\psi \right]\nn\\
&\quad+\dicg\,\left(\theta_\gamma-\theta_\mathrm{DM}\right) + \icg\,\left(\dot{\theta}_\gamma - \dot{\theta}_\mathrm{DM}\right)\nn\\
&\quad +\ha\,\dot{\theta}_b + \ha^2\theta_b - k^2\,\ha\,c_b^2\delta_b + k^2\ha\,\psi - 2\ha\,\dot{\theta}_\gamma\,,
\end{align}
where $\bar{c_b^2} = (\ha c_b^2 + \dot{c_b^2})$. This expression for $\dot{g}_\Theta$ can be further simplified by using Eq.~(\ref{eq: boltzmann-theta_b_dot}) and Eq.~(\ref{eq: boltzmann-h-cg-F1}) together with the relation $\dicg= -2\ha\icg$ to give
\begin{align}
&\dot{g}_\Theta =
2\ha\Tgb - \frac{\ddot{a}}{a}\theta_b + k^2\left[ \bar{c_b^2}\delta_2 + c_b^2\dot{\delta}_b - \frac{1}{4}\dot{\delta}_\gamma +\dot{\sigma}_\gamma  \right.\nn\\
&\quad\left.- \ha\psi -\frac{\ha}{2}\delta_\gamma + 2\ha \sigma_\gamma\right] + \icg\left( \dot{\theta}_\gamma - \dot{\theta}_\mathrm{DM}\right)\nn\\
&\quad+ \ha\,\kd\, (2+R)\, \Theta_{\gamma b}\,.
\end{align}
Note that for a temperature independent scattering cross section the occurrence of the term proportional to $\icg$ in the evolution of $\theta_\gamma$ cancels out the $\dicg(\theta_\gamma - \theta_\mathrm{DM})$ term from $\dot{g}_\Theta$. The first order expression for the photon-baryon slip in the presence of dark-matter photon interactions is now given by
\begin{align}
&\Tgb = \left(\frac{\dot{\tc}}{\tc} -\frac{2\ha}{1+R} \right)\Theta_{\gamma b} - \frac{\tc}{1+R}\left[-\frac{\ddot{a}}{a}\theta_b + k^2\left(\bar{c_b^2}\delta_b \right.\right.\nn\\
&\quad\left.\left.+ c_b^2\dot{\delta}_b - \frac{1}{4}\dot{\delta}_\gamma - \frac{\ha}{2}\delta_\gamma - \ha\psi \right) + \icg\left(\dot{\theta}_\gamma - \dot{\theta_\mathrm{DM}} \right) \right]  + \mathcal{O}(\tc^2)\,.
\label{eq: tca-icg-slip-o1}
\end{align}
Similar expressions have been obtained for scenarios, in which the tight coupling regime is modified by a coupling term between dark matter and baryons \cite{Sigurdson:2004zp,Chen:2002yh}. Extending the calculation to second order in $\tc$, we obtain
\begin{align}
\Tgb &= \left(1 - 2\ha\frac{\tc}{1+R}\right)\left\lbrace\left( \frac{\dot{\tc}}{\tc} - \frac{2\ha}{1+R} \right)\Theta_{\gamma b} - \frac{\tc}{1+R}\times\right.\nn\\
&\left.\times\left[ -\frac{\ddot{a}}{a}\theta_b + k^2\left( \bar{c_b^2}\delta_b + c_b^2\dot{\delta}_b -\frac{1}{4}\dot{\delta}_\gamma - \frac{\ha}{2}\delta_\gamma - \ha\psi\right)\right.\right.\nn\\
&\left.\left. + \icg\left( \dot{\theta}_\gamma - \dot{\theta}_\mathrm{DM} \right) \right] \right\rbrace-\frac{\tc}{1+R}\,k^2\left(2\ha\,\sigma_\gamma + \dot{\sigma}_\gamma \right) \nn\\
& + \tilde{f}_\Theta\left(\ddot{\tilde{f}}_\Theta g_\Theta + 2\dot{\tilde{f}}_\Theta\dot{g}_\Theta + \tilde{f}_\Theta\ddot{g}_\Theta \right) + \mathcal{O}(\tc^3)\,.
\end{align}

To solve the evolution of $\Tgb$ at second order in $\tc$ a first order expression for $\sigma_\gamma$ is needed, while the evolution of $\dot{\theta}_\gamma$ requires knowledge of $\sigma_\gamma$ at order $\tc^2$. As we will show shortly, all further multipoles contribute at sub-dominant order and can be neglected during the tight coupling regime. Polarisation multipoles at $l=0,2$ obey (cf. Eq.~(\ref{eq: boltzmann-h-gcdm-photons}))
\begin{subequations}
\begin{align}
&-\frac{1}{2}G_{\gamma 0} + \sigma_\gamma + \frac{1}{2}G_{\gamma 2} = \tau_2\left[\dot{G}_{\gamma 0} + k G_{\gamma 1} \right]\,,\\
&-G_{\gamma 2} + \frac{1}{10}\left(2\sigma_\gamma + G_{\gamma 0} + G_{\gamma 2}\right)\nn\\
&\quad= \tau_2\left[\dot{G}_{\gamma 2} - \frac{k}{5}\left(2G_{\gamma 1} - 3 G_{\gamma 3} \right) \right]\,,
\end{align}
\label{eq: tca-shear-G02-reformulation}
\end{subequations}
where we abbreviated $\tau_2 \equiv \left(\dot{\kappa}+\icg \right)^{-1}$. Because $\icg\ll\kd$ $\tau_2\sim\mathcal{O}(\tc)$, and we see from Eq.~(\ref{eq: tca-shear-G02-reformulation}) that at leading order in $\tc$ $G_{\gamma0}\sim G_{\gamma}\sim\sigma_\gamma\sim\mathcal{O}(\tc)$. The leading order behaviour of the higher order multipoles is given by
\begin{equation}
F_{\gamma l} = \frac{l\,k\,\tau_2}{2l+1}\,F_{\gamma(l-1)}\,, \quad \mathrm{and}\quad G_{\gamma l} = \frac{l\,k\,\tau_2}{2l+1}\,G_{\gamma(l-1)}\,,
\end{equation}
while for the polarisation dipole we obtain
\begin{equation}
G_{\gamma 1} = \tau_2\left[ -\dot{G}_{\gamma 1} + \frac{k}{3}\left(G_{\gamma 0} - 2 G_{\gamma 2} \right) \right]\,.
\end{equation}
Thus $F_{\gamma3}\sim G_{\gamma1} \sim G_{\gamma3} \sim \mathcal{O}(\tc^2)$ and all multipoles $l\ge4$ only contribute at higher orders in $\tc$. From Eq.~(\ref{eq: boltzmann-h-cg-F2}) we find
\begin{equation}
\sigma_\gamma = \frac{8\tau_2}{27}\theta_\gamma + \frac{1}{18} \left(G_{\gamma 0} + G_{\gamma 2} \right) - \frac{10\tau_2}{9}\dot{\sigma}_\gamma + \mathcal{O}(\tc^3)\,.
\label{eq: tca-shear-2}
\end{equation}
We can use Eq.~(\ref{eq: boltzmann-h-cg-G0}) and Eq.~(\ref{eq: boltzmann-h-cg-G2}) to obtain approximate expressions for the photon polarisation
\begin{subequations}
\begin{align}
G_{\gamma 0} &= -2\tau_2\dot{G}_{\gamma 0} + 2\sigma_\gamma + G_{\gamma 2} + \mathcal{O}(\tc^3)\,\nn\\
&=\frac{5}{2}\sigma_\gamma - \frac{25}{4}\tau_2\dot{\sigma}_\gamma  + \mathcal{O}(\tc^3)\,,\\
G_{\gamma 2} &= \frac{10\tau_2}{9}\left(\frac{2k}{5} G_{\gamma 1} -\dot{G}_{\gamma 2} \right) + \frac{2}{9}\sigma_\gamma + \frac{1}{9}G_{\gamma 0}+ \mathcal{O}(\tc^3)\nn\\
&=\frac{1}{2}\sigma_\gamma - \frac{5}{3}\tau_2\dot{\sigma}_\gamma+ \mathcal{O}(\tc^3)\,.
\label{eq: tca-G2-1}
\end{align}
\end{subequations}
These expressions are also used to give initial conditions for the integration of the full Boltzmann hierarchy (c.f. Eq.~(\ref{eq: boltzmann-h-gcdm-cdm})), once the approximation of tight coupling loses it's validity. We find for the photon shear at first order
\begin{equation}
\sigma_\gamma = \frac{16}{45}\tau_2\theta_\gamma + \mathcal{O}(\tau_2^2)\,,
\label{eq: tca-icg-sigma-o1}
\end{equation}
and to second order we obtain
\begin{equation}
\sigma_\gamma = \frac{8\tau_2}{45}\left[ 2\theta_\gamma\left(1-\frac{11}{6}\dot{\tau_2} \right) - 2\dot{\theta}_\gamma\left(\frac{11}{6}\tau_2 \right) \right] + \mathcal{O}(\tau_2^3)\,.
\label{eq: tca-icg-sigma-o2}
\end{equation}

Finally, a comment on the validity of the tight coupling approximation is in order. We identified three conditions, $\tc\,\ha\ll1$, $\tc\,k\ll1$, and $\tc\,\icg\ll1$, of which the former two are also present in $\Lambda$CDM. For the latter we find in the early universe, before the epoch of recombination
\begin{equation}
\tc\icg = \ugc\, \frac{\rho_\mathrm{DM}}{100\,\mathrm{GeV}}\,\frac{1}{n_e}
\simeq 10^{-2}~\ugc~ \frac{\Omega_{\mathrm{DM},0}}{\Omega_{b,0}}\,.
\end{equation}
The critical values of $\tc\,\ha$ and $\tc\,k$, that determine when the tight coupling approximation is no longer valid, and one needs to integrate the full Boltzmann equations, are larger than $10^{-3}$. Therefore the additional requirement on $\tc\icg$ is automatically satisfied in all scenarios with realistic cosmological parameters as long as $\ugc\lesssim 0.01$. 

\subsection{Impact on CMB spectra}
\label{sec: impact}
\begin{figure}
\includegraphics[]{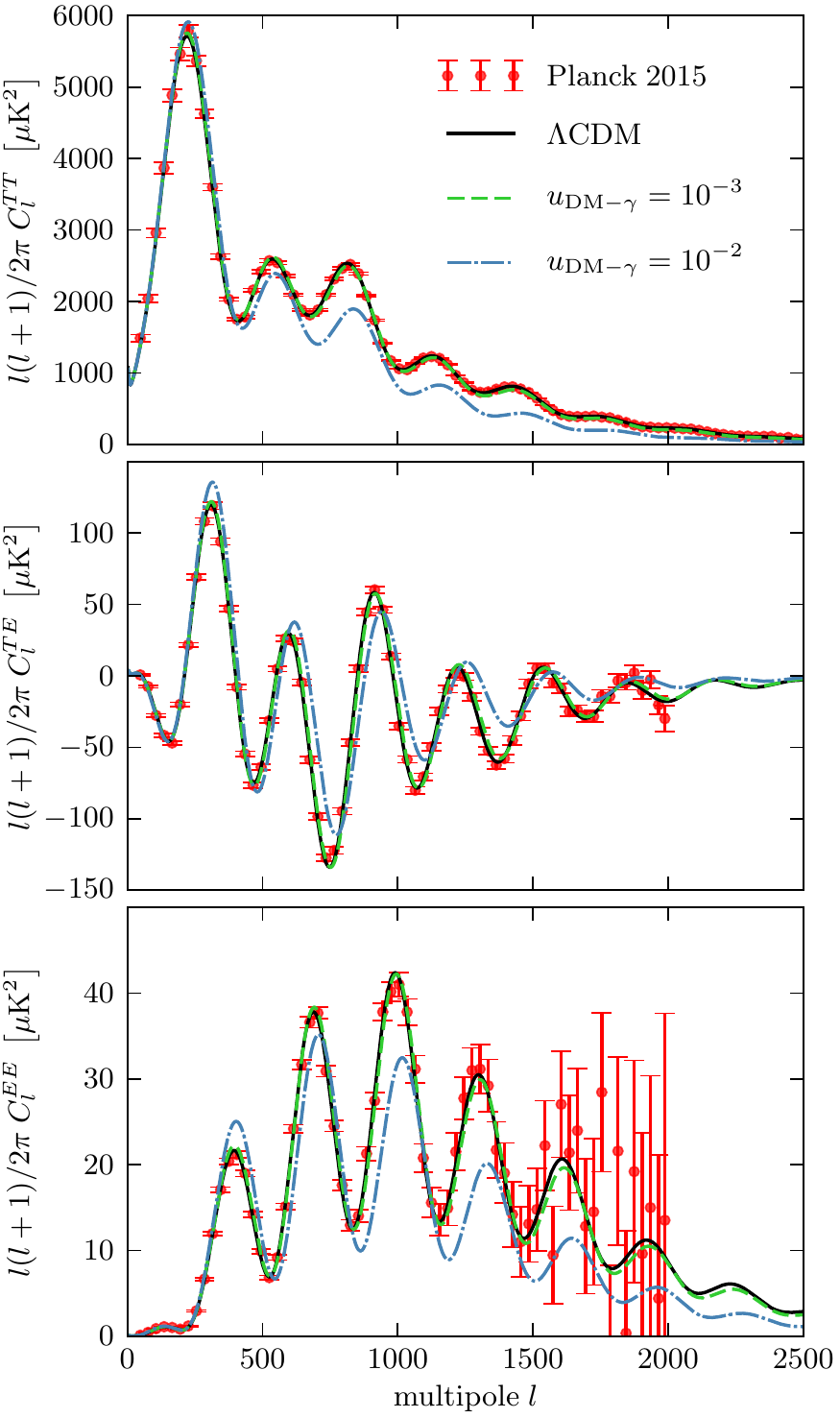}
\caption{The temperature, E-mode polarisation, and TE cross correlation CMB angular power spectra computed from Planck best-fit parameters ($\Lambda$CDM) and in the presence of a non-zero dark matter-photon scattering cross section. Red data points show the errors bars associated with the Planck best fit model.}
\label{fig: Cls}
\end{figure}

The effects of dark matter-photon scattering on the CMB temperature and polarisation spectra have been discussed in Ref.~\cite{Boehm:2001hm, Wilkinson:2013kia} and are shown in Fig.~\ref{fig: Cls}. There are three major effects: (a) the reduction in magnitude of the acoustic peaks at small scales by collisional damping, (b) a shift in the position of the largest Doppler peak towards higher multipoles caused by the decreased sound speed of the photon plasma, and (c) the enhancement of the first acoustic peaks due to a decrease in the photon's diffusion length.

In this section we quantify the impact, which the inclusion of the dark matter sound speed and our modifications to the tight coupling regime have on the computed CMB spectra.

\subsubsection{Effect of the dark matter sound speed}
\label{sec: impact-soundspeed}

\begin{figure}
\includegraphics[]{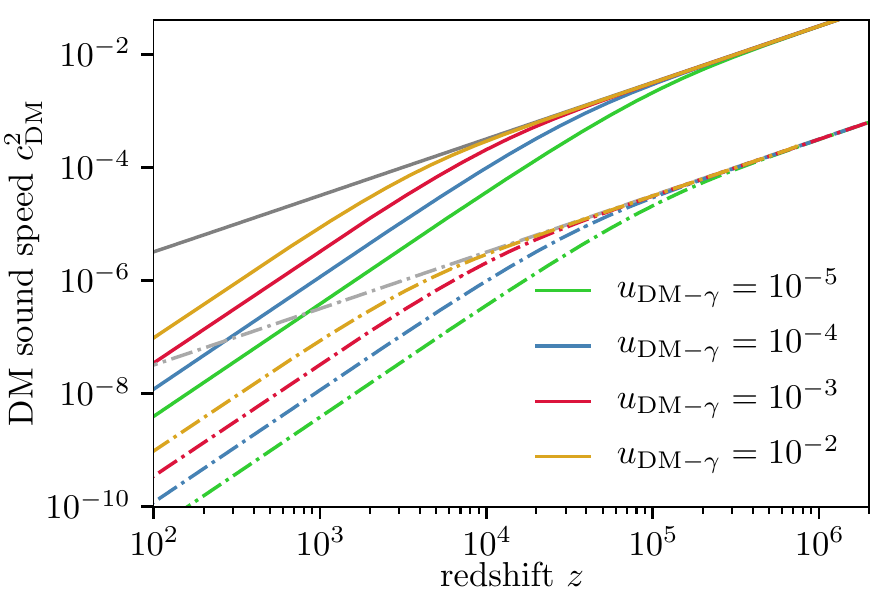}
\caption{The dark matter sound speed for a dark matter mass of $m_\mathrm{DM} = 10~\mathrm{keV}$ (solid) and $m_\mathrm{DM} = 1~\mathrm{MeV}$ (dashed). Shown in grey is the dark matter sound speed if the dark matter and photons maintained equilibrium (i.e. $T_\gamma = T_\mathrm{DM}$) throughout their whole evolution.}
\label{fig: soundspeed}
\end{figure}

The dark matter sound speed evolves in two stages, which can be clearly distinguished in Fig.~\ref{fig: soundspeed}: when the interaction rate between the dark matter and the photons is large enough to maintain both species in kinetic equilibrium, the dark matter temperature is equal to the photon temperature and the sound speed evolves as $a^{-1}$. Once these interactions cease to be effective the dark matter temperature redshifts as $a^{-2}$ and the sound speed decreases faster in this regime. The cross section to mass ratio $\ugc$ sets the time of decoupling and hence has some small impact on the value of $c_\mathrm{DM}$. However, the sound speed is inversely proportional to the dark matter mass (c.f. Eq.~\ref{sound_speed}) and  this parameter governs the magnitude and relevance of $c_\mathrm{DM}$.

To quantify the impact of the dark matter sound speed on our results, we compare a set of CMB spectra for which the sound speed was arbitrarily set to zero to some spectra obtained with the same $\Lambda$CDM $+~\ugc$ parameter values but different dark matter masses. The relevant spectra for the comparison with the Planck measurements are the temperature auto-correlation (TT), the E-mode-polarisation auto correlation (EE) and the temperature E-mode cross correlation (TE) spectrum. As expected, the differences depend only mildly on $\ugc$ but strongly on the dark matter mass. The differences are smaller when the dark matter-photon coupling is smaller and are larger as the dark matter mass decreases. Tab.~\ref{tab: soundspeed-effect} lists the maximum differences obtained for a cross section to mass ratio $\ugc = 0.02$. For all the three spectra, we find that -- for a dark matter mass of at least $10~\mathrm{keV}$ -- the inclusion of the dark matter sound speed affects our computations below the percent level. The maximum effect on the TT-spectrum was found to be $0.3~\%$ and on the EE-spectrum $0.8~\%$. Those masses, which are small enough for the dark matter sound speed to have a sizeable effect on the CMB spectra, imply a sound speed close to unity at the beginning of the evolution of the smallest modes and are not subject to this work.

We will therefore neglect the dark matter sound speed for the reminder of this work as this has the advantage of removing the dark matter mass and reducing the number of additional parameters.
\renewcommand{\arraystretch}{1.2}
\begin{table}
\begin{center}
\begin{tabular}{L{0.07\textwidth}|C{0.1\textwidth}|C{0.1\textwidth}|C{0.1\textwidth}}
$m_\mathrm{DM}$ & $\Delta C_l^{TT}$ & $\Delta C_l^{TE}$ & $\Delta C_l^{EE}$ \\
\hline
$10~\mathrm{keV}$ & $13.4 ~\upmu\mathrm{K}^2$ & $0.74  ~\upmu\mathrm{K}^2$ & $0.08~\upmu\mathrm{K}^2$ \\
$1~\mathrm{MeV}$  & $0.14 ~\upmu\mathrm{K}^2$ & $0.007~\upmu\mathrm{K}^2$ & $8\times10^{-4}~\upmu\mathrm{K}^2$ \\
$1~\mathrm{GeV}$  & $0.003~\upmu\mathrm{K}^2$ & $0.0003~\upmu\mathrm{K}^2$ & $4\times10^{-5}~\upmu\mathrm{K}^2$
\end{tabular}
\caption{Maximum values of the effect of the dark matter sound speed on the CMB spectra, computed for $\ugc = 0.01$.}
\label{tab: soundspeed-effect}
\end{center}
\end{table}

\begin{figure}
	\includegraphics[]{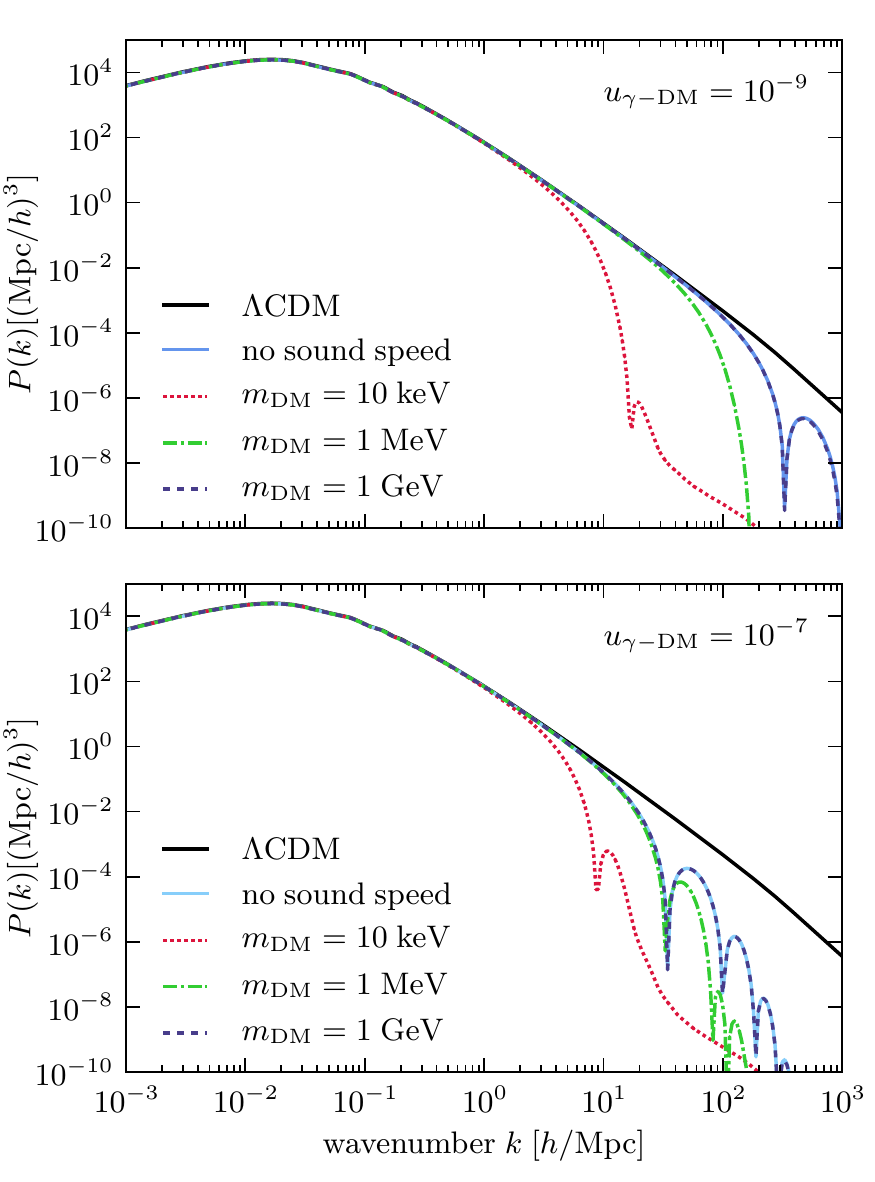}
	\caption{Effect of the dark matter sound speed on the matter power spectrum at small scales for different cross section to mass ratios and dark matter masses.}
	\label{fig: soundspeed-pk}
\end{figure}

Indeed, the scales probed by CMB experiments are too large for the dark matter sound speed to have a sizeable effect. We note, however, that there may be an impact at smaller scales. These can be probed by N-body simulations of structure formation \cite{Boehm:2014vja,Schewtschenko:2014fca,Schewtschenko:2015rno,Escudero:2015yka}, which require the linear matter power spectrum as an input. In Fig. \ref{fig: soundspeed-pk} we show the matter power spectrum obtained with different values of $m_\mathrm{DM}$ for a scattering cross section to mass ratio of $\ugc=10^{-7}$ and $\ugc=10^{-9}$. The dark matter sound speed leads to a suppression of small scale structure on its own.
As the magnitude of $c_\mathrm{DM}$ is dominantly set by the dark matter mass, this parameter also determines the scale at which the suppression occurs due to the sound speed. On the other hand, the scale at which the suppression of structure caused by the scattering terms in the Boltzmann equations becomes becomes relevant is entirely determined by $\ugc$. In particular for small dark matter masses the former effect can dominate, i.e. the sound speed causes suppression at larger scales than the scattering terms.
In this case, the matter power spectra computed with and without taking into account sound speed contributions differ from each other considerably.
From Fig.~\ref{fig: soundspeed-pk} it is evident that, for the smallest dark matter masses considered here, the damping scale depends on $\ugc$ only very mildly. As the dark matter mass increases, the suppression of power induced by the collisional damping effect starts to dominate and $c_\mathrm{DM}$ can be safely neglected. The larger the value of $\ugc$, 
the smaller the dark matter mass at the transition point. We have checked that for cross sections to mass ratios as small as $\ugc=10^{-9}$ the zero sound speed result is recovered for dark matter masses of $1~\mathrm{GeV}$ (c.f. Fig.~\ref{fig: soundspeed-pk}). In light of these findings the results of previous studies on the non-linear evolution of dark matter scattering with photons \cite{Boehm:2014vja,Schewtschenko:2014fca,Schewtschenko:2015rno,Escudero:2015yka} remain valid under the assumption of a large enough dark matter mass. For smaller dark matter masses one can potentially derive stronger constraints by taking into account the dark matter sound speed correctly.

\subsubsection{Impact of modification to the tight coupling approximation}
\label{sec: impact-tca}
\begin{figure}
\includegraphics[]{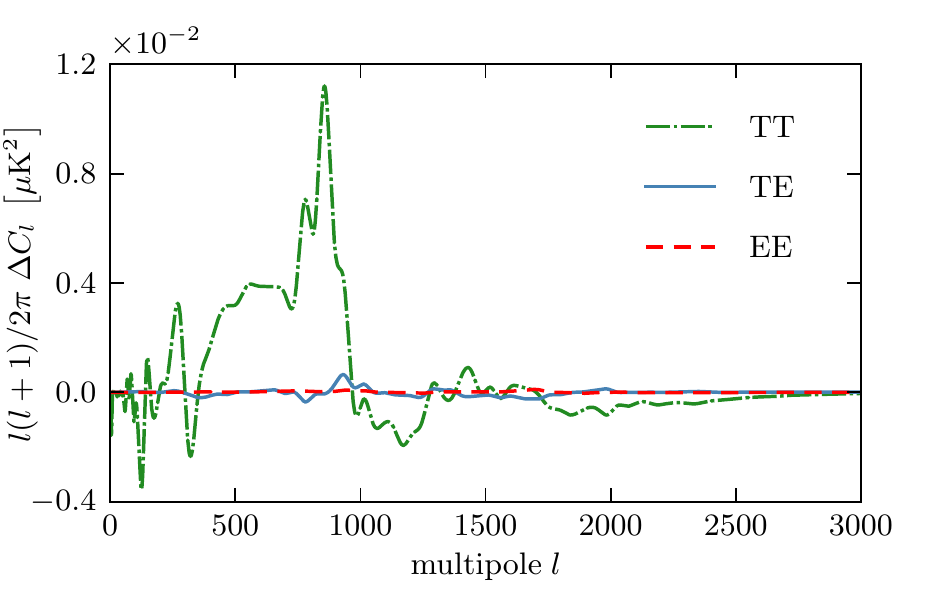}
\caption{Difference between the CMB temperature and polarisation spectra computed from the expressions found in section \ref{sec: equations} and those used in Ref.~\cite{Wilkinson:2013kia}. The plot was obtained for a scattering cross section to mass ratio of $\ugc=10^{-4}$.}
\label{fig: differences}
\end{figure}

To compare our results, obtained for zero dark matter sound speed, with those of previous works we ran the same code as was used in Ref.~\cite{Wilkinson:2013kia}. The main discrepancy between our work and previous approximations is a slightly different expression for the tight coupling approximation (see Eq.~(\ref{eq: tca-icg-slip-o1})). Comparing the CMB angular power spectra obtained with both codes, we find that the largest differences occur for the temperature spectrum and can reach up to $10~\upmu\mathrm{K}^2$. However, the code used in Ref.~\cite{Wilkinson:2013kia} is based on CLASS version 1.6, and CLASS itself has undergone major changes since then \cite{Lesgourgues:2013bra}. Moreover, the default values of many cosmological and precision parameters in CLASS, such as e.g. the parameters describing reionization or the primordial helium abundance during BBN, have changed, and, for a meaningful comparison, they need to be set to the same value in all codes by hand. To determine the importance of the tight coupling regime, we transferred the code used in Ref.~\cite{Wilkinson:2013kia} to an up-to-date version of CLASS. The resulting differences are depicted in Fig.~\ref{fig: differences} and are most prominent for the temperature cross-correlation spectrum. However, even in this case, they appear to be smaller than $0.01~\upmu\mathrm{K}^2$, which is well below the experimental sensitivity. We therefore conclude that the results of Ref.~\cite{Wilkinson:2013kia} are robust against our modifications.

\section{Parameter Extraction}
\label{sec: parameter-extraction}
\subsection{Methodology}
\begin{table*}[ht]
\renewcommand{\arraystretch}{1.2}
\centering
\begin{tabular}{L{.14\textwidth}|C{.18\textwidth}|C{.18\textwidth}|C{.18\textwidth}|C{.18\textwidth}}
& \specialcell{Planck TT + lowTEB\\[-3pt] ~} &
  \specialcell{Planck TT + lowTEB\\[-3pt] + lensing} &
  \specialcell{Planck TT + lowTEB\\[-3pt] + $N_\mathrm{eff}$} & 
  \specialcell{Planck TT + lowTEB\\[-3pt] + lensing + $N_\mathrm{eff}$} \\
 
\hline\\[-22pt] & & & & \\
$\Omega{_{b}{h^2}}$ &
$0.02231_{-0.00025}^{+0.00022}$ & 
$0.0223_{-0.00024}^{+0.00023}$&
$0.02238_{-0.00039}^{+0.00036}$ &
$0.0224_{-0.00039}^{+0.00034}$ 
\\

$\Omega{_{c}{h^2} }$ &
$0.1198\pm0.0023$ &
$0.119\pm0.002$&
$0.1208_{-0.0043}^{+0.0039}$ &
$0.1203\pm0.004$ 
\\

$H_{0 }\,\left[\mathrm{km}/(\mathrm{Mpc\,s}) \right]$ &
$67.56_{-1.00}^{+0.94}$ &
$67.83_{-0.93}^{+0.91}$ &
$68.27_{-2.9}^{+2.6}$ &
$68.78_{-3.0}^{+2.5}$ 
\\ 

$\ln\left(10^{10}A_{s}\right)$ &
$3.094_{-0.038}^{+0.033}$ & 
$3.072_{-0.030}^{+0.028}$&
$3.1_{-0.050}^{+0.043}$ &
$3.082_{-0.045}^{+0.033}$ 
\\

$n_{s }$ &
$0.9661_{-0.0064}^{+0.0061}$ & 
$0.9672_{-0.0063}^{+0.0055}$ &
$0.9696_{-0.017}^{+0.014}$ &
$0.9721_{-0.016}^{+0.015}$ 
\\

$\tau_{reio }$ &
$0.08005_{-0.019}^{+0.018}$ & 
$0.07022_{-0.017}^{+0.016}$ &
$0.08205_{-0.025}^{+0.018}$ &
$0.07405_{-0.022}^{+0.015}$
\\

$10^{+4}~u_{{\gamma-c} }$ &
$< 2.254$ &
$< 1.805$ &
$< 2.142$ &
$< 1.818$
\\ 

$N_\mathrm{eff}$ &
$3.046$ &
$3.046$ & 
$3.137_{-0.35}^{+0.29}$ &
$3.161_{-0.33}^{+0.30}$
\\

$\sigma_{8 }$ &
$0.8003_{-0.02}^{+0.029}$ & 
$0.7942_{-0.014}^{+0.024}$ &
$0.8057_{-0.028}^{+0.032}$ &
$0.7999_{-0.023}^{+0.028}$ \\
 
\end{tabular}
\caption{Cosmological parameters found from datasets without the high multipole polarisation data when allowing for dark matter-photon scattering. Upper and lower bounds correspond to the 68\% CL interval, when only upper limits are shown they correspond to 95\% c.l. limits.}
\label{tab: PlanckTT+lowTEB}
\end{table*}

We use our modified version of CLASS v2.6 together with the Monte Carlo code MontePython\footnote{http://baudren.github.io/montepython.html} \cite{Audren:2012wb} to effectively sample the model's parameter space, following the methodology of Ref.~\cite{Wilkinson:2013kia,DiValentino:2017oaw}. The latter corresponds to the six parameters of the standard $\Lambda$CDM model, namely the baryon density $\Omega_b h^2$, the dark matter density $\Omega_\mathrm{DM} h^2$, the ratio between the sound horizon and the angular diameter distance at decoupling $\theta_s$, the reionization optical depth $\tau_\mathrm{reio}$, and the amplitude $A_s$ and spectral index $n_s$ of primordial perturbations plus the dark matter-photon coupling strength $\ugc$.

In this baseline scenario we assume two massless and one massive neutrino of $m_\nu=0.06~\mathrm{eV}$ and adjust the effective number of neutrinos to the standard model prediction of $N_\mathrm{eff}=3.046$ \cite{Mangano:2005cc}. In an additional run we also let the effective number of radiative degrees of freedom $N_\mathrm{eff}$ free to vary.

To understand how the polarisation data affects our conclusions, we start by analysing the Planck 2015 temperature data over the full range of multipoles ($2 \le l \le 2508$) plus the polarisation data at low multipoles ($2 \le l \le 29$) \cite{Aghanim:2015xee}. We refer to this dataset as ``Planck TT + lowTEB". We then include the polarisation data at high multipoles up to $l=1996$ and refer to this set as ``Planck TTTEEE + lowTEB". The Planck collaboration considers this data set as preliminary due to unresolved systematic at the level of $\mathcal{O}(1)~\upmu\mathrm{K}^2$ \cite{Aghanim:2015xee}. Finally we extend both datasets by the 2015 Planck measurements of the CMB lensing potential power spectrum $C_l^{\phi\phi}$ \cite{Ade:2015zua}. We refer to these data sets as "+ lensing".

All scenarios, for which the inferred value of $H_0$ is in less than $2\sigma$ tensions with local measurements from Cepheids variables \cite{Riess:2016jrr} are reanalysed. For that we assume a Gaussian prior on $H_0$ ($H_0 = 73.24\pm1.75~\mathrm{km}(\mathrm{s~Mpc})^{-1}$), by opposition to the flat prior used to analyse previous datasets. We refer to this analysis by the extension "+ R16".

In addition to the model parameters mentioned above the Planck temperature likelihood at high multipoles contains 15 nuisance parameters (and further 12 if polarisation data is considered), modelling foreground effects \cite{Aghanim:2015xee}. These are sampled jointly with the model parameters and then marginalised over to derive our final results. To check the correctness of our parameter extraction we run $\Lambda$CDM test scenarios for several combinations of data sets. 

For these test sets we find that some of the nuisance parameters differ by roughly $1\sigma$ from the results published by the Planck collaboration\footnote{http://wiki.cosmos.esa.int/planckpla2015/images/f/f7/\newline Baseline\_params\_table\_2015\_limit68.pdf}. We believe this is connected to the nuisance parameters $\mathcal{D}^\mathrm{kSZ}$ and $\mathcal{D}^{tSZ}$, which describe the amplitude of the kinetic (kSZ) and the thermal (tSZ) Sunyaev–Zel'dovich effect, respectively. While the Planck collaboration employs an additional joint Gaussian prior $\mathcal{D}^\mathrm{kSZ} + 1.6\times\mathcal{D}^{tSZ} = \left(9.5\pm0.3\right)~\upmu\mathrm{K}^2$ \cite{Ade:2015xua}, only the individual flat priors on both parameters are implemented in MontePython. The purpose of this additional Gaussian prior, used by the Planck collaboration, is to keep the tSZ and kSZ amplitudes from drifting into unphysical regions of the parameter space, which have been previously excluded by the SPT experiment \cite{Reichardt:2011yv}. 

Ultimately, we are only interested in the correct determination of the model parameters, for which the central values vary by $0.4\sigma$ with respect to Planck results in the most extreme cases. Comparable shifts have been obtained by the Planck collaboration, comparing results obtained with the codes CAMB and PICO \cite{Aghanim:2015xee}. To further verify the validity of our results, we use the ``lite" likelihood, which has been marginalised over nuisance parameters and thus is unaffected by the issue described above. We find a very good agreement between the confidence intervals for cosmological parameters extracted from the ``lite" and the full likelihood. Therefore we believe the intervals found for our model parameters are accurate.

\subsection{Results based on the "Planck TT + lowTEB" data set}
\label{sesec: res-PlanckTT+lowTEB}
\begin{figure}
\includegraphics[]{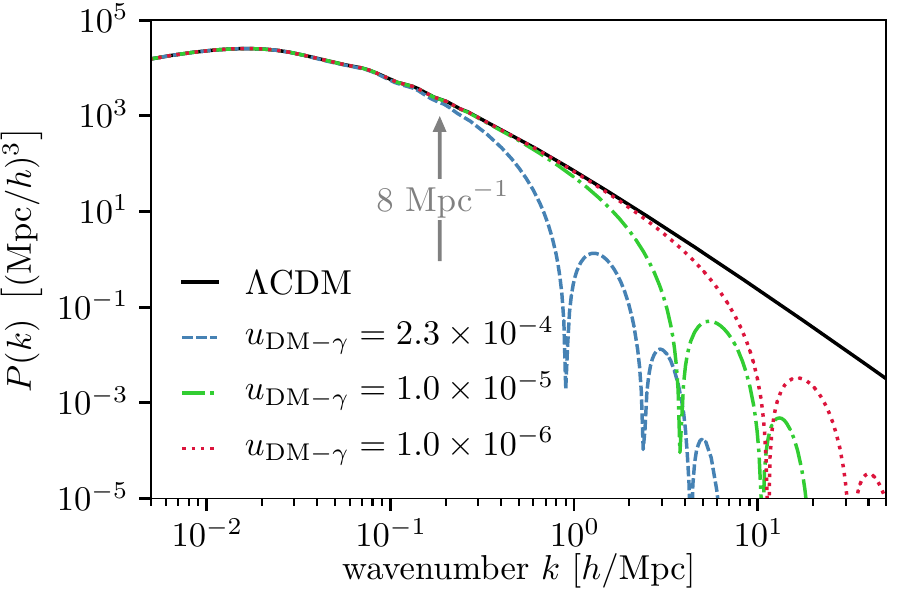}
\caption{Linear matter power spectrum in the presence of dark matter-photon scattering. This plot was obtained assuming a large mass for the dark matter particle, i.e. neglecting the dark matter sound speed.}
\label{fig: pk}
\end{figure}

Results obtained from datasets without the high multipole polarisation data are given in Tab.~\ref{tab: PlanckTT+lowTEB}. Using the ``Planck TT + lowTEB" dataset, we find a 95\% CL upper limit on the dark matter-photon scattering cross section $\sigma_{\mathrm{DM}-\gamma} \leq 2.25 \times 10^{-6}~ \sigma_\mathrm{Th} \left(m_\mathrm{DM}/\mathrm{GeV}\right)$. This limit is about 20\% stronger than those derived in Ref.~\cite{Wilkinson:2013kia}, using the 2013 data release. The constraint tightens up by another 20\% when the lensing data is included in the analysis.

We also find that the limits on $\ugc$ are only marginally affected when the effective number of neutrinos is set as a free parameter. While the central values found for $N_\mathrm{eff}$ lie somewhat above the standard model prediction of $N_\mathrm{eff}=3.046$, these discrepancies remain well below $1\sigma$ for all scenarios.

For the scenarios where $N_\mathrm{eff}$ is free to vary, we observe that the value of $H_0$ increases slightly. This increase can be caused by two distinct effects. Firstly, interactions between dark matter and photons increase the magnitude of the first acoustic CMB peaks. This increase is less pronounced in a younger universe, hence the larger $H_0$ value. Secondly, there is a well-known degeneracy between $H_0$ and $N_\mathrm{eff}$ that leads to larger values of $H_0$ when the effective number of neutrinos is free to vary.

\begin{table*}
\renewcommand{\arraystretch}{1.2}
\centering
\begin{tabular}{L{.14\textwidth}|C{.18\textwidth}|C{.18\textwidth}|C{.18\textwidth}|C{.18\textwidth}}
& \specialcell{Planck TTTEEE\\[-3pt] + lowTEB\\[-3pt]~} & 
  \specialcell{Planck TTTEEE\\[-3pt] + lowTEB\\[-3pt] + lensing} &
  \specialcell{Planck TTTEEE\\[-3pt] + lowTEB\\[-3pt] + $N_\mathrm{eff}$} & 
  \specialcell{Planck TTTEEE\\[-3pt] + lowTEB\\[-3pt] + lensing + $N_\mathrm{eff}$} \\
 
\hline\\[-22pt] & & & & \\
$\Omega{_{b}{h^2}}$ &
$0.02228\pm0.00016$ &   
$0.02228_{-0.00017}^{+0.00015}$ &
$0.02221_{-0.00026}^{+0.00022}$ &
$0.02218_{-0.00024}^{+0.00022}$ \\

$\Omega{_{c}{h^2} }$ &
$0.1201_{-0.0016}^{+0.0014}$ &
$0.1197_{-0.0014}^{+0.0015}$ &
$0.1192_{-0.0032}^{+0.0030}$ &
$0.1182_{-0.0028}^{+0.0031}$ \\

$H_{0 }\,\left[\mathrm{km}/(\mathrm{Mpc\,s}) \right]$ &
$67.33_{-0.66}^{+0.67}$ &
$67.52_{-0.60}^{+0.66}$ &
$66.8\pm1.6$ &
$66.78_{-1.6}^{+1.5}$ \\ 

$\ln\left(10^{10}A_{s}\right)$ &
$3.097_{-0.028}^{+0.038}$ &
$3.069_{-0.028}^{+0.021}$ &
$3.087_{-0.039}^{+0.035}$ &
$3.06_{-0.028}^{+0.026}$ \\

$n_{s }$ &
$0.964_{-0.0048}^{+0.0046}$ &
$0.9646_{-0.0047}^{+0.0051}$ &
$0.961_{-0.0095}^{+0.0096}$ &
$0.9604_{-0.0095}^{+0.0091}$ \\

$\tau_{reio }$ &
$0.08107_{-0.016}^{+0.019}$ &
$0.06755_{-0.014}^{+0.012}$ &
$0.07734_{-0.018}^{+0.016}$ &
$0.06557_{-0.014}^{+0.012}$ \\

$10^{+4}~u_{{\gamma-c} }$ &
$< 1.579$ &
$< 1.490$ &
$< 1.623$ &
$< 1.359$ \\ 

$N_\mathrm{eff}$ &
$3.046$ &
$3.046$ &
$2.974_{-0.21}^{+0.20}$ &
$2.943_{-0.20}^{+0.19}$ \\

$\sigma_{8 }$ &
$0.8103_{-0.018}^{+0.024}$ &
$0.7982_{-0.012}^{+0.022}$ &
$0.8036_{-0.021}^{+0.027}$ &
$0.7946_{-0.016}^{+0.021}$ \\ 
 
\end{tabular}
\caption{Cosmological parameters found from datasets including temperature and polarisation data at low and high multipoles when allowing for dark matter-photon scattering. Upper and lower bounds correspond to the 68\% CL interval, when only upper limits are shown they correspond to 95\% c.l. limits.}
\label{tab: PlanckTTTEEE+lowTEB}
\end{table*}
\begin{table}
\centering
\renewcommand{\arraystretch}{1.2}
\begin{tabular}{l|C{.16\textwidth}|C{.16\textwidth}}
& \specialcell{ Planck TT \\[-3pt]+ low TEB + $N_\mathrm{eff}$} &
  \specialcell{ Planck TT \\[-5pt] + low TEB \\[-5pt] + lensing + $N_\mathrm{eff}$} \\
  
\hline\\[-22pt] & & \\
$\Omega{_{b}{h^2}}$ &
$0.02241_{-0.00039}^{+0.00036}$ &
$0.02246_{-0.0046}^{+0.0031}$
\\

$\Omega{_{c}{h^2} }$ &
$0.1212_{-0.0044}^{+0.0034}$ &
$0.1208\pm0.0039$
\\

$H_{0 }\,\left[\mathrm{km}/(\mathrm{Mpc\,s}) \right]$ &
$68.36_{-3.0}^{+2.6}$ &
$69.27_{-3.5}^{+2.3}$
\\ 

$\ln\left(10^{10}A_{s}\right)$ &
$3.104_{-0.05}^{+0.04}$ &
$3.089_{-0.047}^{+0.037}$
\\

$n_{s }$ &
$0.9704_{-0.017}^{+0.014}$ &
$0.9746_{-0.02}^{+0.012}$
\\

$\tau_{reio }$ &
$0.08369_{-0.025}^{+0.018}$ &
$0.07686_{-0.021}^{+0.017}$
\\

$10^{+4}~u_{{\gamma-c} }$ &
$< 2.488$ &
$< 1.724$
\\ 

$N_\mathrm{eff}$ &
$3.154_{-0.35}^{+0.27}$ &
$3.215_{-0.39}^{+0.25}$
\\

$\sigma_{8 }$ &
$0.805_{-0.025}^{+0.035}$ &
$0.8048_{-0.025}^{+0.027}$
\\ 

\end{tabular}
\caption{Cosmological parameters found when including the R16 dataset. Upper and lower bounds correspond to the 68\% CL interval, when only upper limits are shown they correspond to 95\% c.l. limits.}
\end{table}

The Planck collaboration reported $H_0 = 68.0^{+2.6}_{-3.0}~\mathrm{km}/(\mathrm{s~Mpc})$ for the $\Lambda$CDM scenario when analysing the "Planck TT + lowTEB + $N_\mathrm{eff}$" datasets and $H_0 = 68.5^{+2.5}_{-3.0}~\mathrm{km}/(\mathrm{s~Mpc})$ by adding the lensing data. These results are very similar to the ones obtained by our analysis (see Tab.~\ref{tab: PlanckTT+lowTEB}), indicating that $H_0$ is well constrained from the measurement of the CMB spectrum and cannot be significantly altered by the inclusion of dark matter-photon scattering. We find that for those datasets where $N_\mathrm{eff}$ is allowed to vary the tension with local measurements of $H_0$ (i.e $H_0 = 73.24\pm 0.75$ \cite{Riess:2016jrr}) is less than $2\sigma$. Therefore we are able to combine these datasets with the R16 data. The results of this analysis are presented in section \ref{sesec: res-R16}.

Finally, we note that all the datasets reported in this section prefer a smaller central value and larger confidence limits for $\sigma_8$ than in the $\Lambda$CDM scenario. The Planck collaboration reported $\sigma_8 = 0.829\pm0.014$ for the "Planck TT + lowTEB" data set and $\sigma_8 = 0.8149\pm0.0093$ if lensing data is included in the analysis. Allowing $N_\mathrm{eff}$ to vary these intervals become $\sigma_8 = 0.834_{-0.025}^{+0.022}$ and $\sigma_8 = 0.820^{+0.018}_{-0.021}$ respectively. In comparison, our value of $\sigma_8$ decreases by roughly 3\% (see Tab.~\ref{tab: PlanckTT+lowTEB}) while the size of the lower confidence interval increases by about 50\% when $N_\mathrm{eff}$ is kept fixed and 10\% if it is free to vary. These combined effects reduce the tensions with the KiDS-450 measurement $S_8 \equiv \sigma_8\sqrt{\Omega_m/0.3} = 0.826^{+0.033}_{-0.028}$ below $1\sigma$. The reason for this change becomes apparent in Fig.~\ref{fig: pk}, where we show the linear matter power spectrum for various interaction strengths. Dark matter-photon interactions lead to a suppression of power on small scales. The largest value of the elastic scattering cross section to mass ratio, that is allowed by CMB observations, lead to a visible suppression of the linear matter power spectrum at 8 Mpc, precisely.

\subsection{Results including high multipole polarisation data}
\label{sesec: res-polarization}
Tab.~\ref{tab: PlanckTTTEEE+lowTEB} shows our results when we include the polarisation data at high multipoles. We note that the addition of information from polarisation at small scales allows to further constrain the $\ugc$ ratio for dark matter-photon interactions. Including data from temperature and polarisation at low and high multipoles as well as lensing potentials, the constraint on the cross section becomes as small as $\sigma_{\mathrm{DM}-\gamma} \leq 1.49\times10^{-6}~\sigma_\mathrm{Th}\left(m_\mathrm{DM}/\mathrm{GeV}\right)$. This makes the constraint 35\% stronger than that derived in Ref.~\cite{Wilkinson:2013kia}.

The addition of small-scale polarisation data partially breaks the degeneracy between $H_0$ and $N_\mathrm{eff}$. Now the data requires values of $H_0$, which are too low to reduce tensions with local measurements below $2\sigma$, even if $N_\mathrm{eff}$ is free to vary. All results we obtain for the effective number of neutrinos are perfectly compatible with the standard model expectation.

As in the previous section, we find lower values for $\sigma_8$ than those obtained for a $\Lambda$CDM-scenario by the Planck collaboration, namely $\sigma_8 = 0.831\pm0.013$ for the "Planck TTTEEE + lowTEB" dataset and $\sigma_8 = 0.8150\pm0.0087$ if lensing data is including. Allowing $N_\mathrm{eff}$ to vary, this confidence limits become $\sigma_8 = 0.828\pm0.018$ and $\sigma_8 = 0.809\pm0.013$. As explained in the previous section, this is due to the onset of collisional damping on the scale of $8~\mathrm{Mpc}$ for the largest couplings allowed by CMB data. Consequently, all datasets presented here are in less than $1\sigma$ tension with the KiDS-450 measurement of $\sigma_8$.

\subsection{Results with R16}
\label{sesec: res-R16}

For those scenarios where $N_\mathrm{eff}$ is let free to vary and polarisation data is ignored, the tension between the value inferred for $H_0$ and local measurements \cite{Riess:2016jrr} is less than $2\sigma$. Applying a Gaussian prior on $H_0$, we, however, notice that larger values of $H_0$ and $N_\mathrm{eff}$ are not supported by CMB data. That is true even when dark matter-photon scattering is allowed. In both cases $H_0$ only shifts very slightly and tensions remain at $1.9\sigma$ and $1.7\sigma$ respectively. Likewise, $N_\mathrm{eff}$ experiences only a small shift towards larger values remaining compatible with the standard model expectation at the $1\sigma$ level.

\section{Conclusions}
\label{sec: conclusions}
In the standard $\Lambda$CDM model dark matter is assumed to be collisionless. However, interacting dark matter models are also compatible with current data. In this work we consider the possibility of elastic scattering between dark matter and photons. Not only would such an effect modify the CMB angular spectra for temperature, polarisation, and lensing, but it would also have a noticeable impact on the matter power spectrum. 

As shown in Fig.~\ref{fig: pk} dark matter-photon interactions introduce a characteristic cut-off scale in the matter power spectrum followed by a series of damped oscillations. This suppression of power can be very significant. For the maximally allowed values of $\ugc$ found in this work, the $P(k)$ would be reduced by roughly two orders of magnitude with respect to $\Lambda$CDM at $k\sim 1 \, h/\mathrm{Mpc}$. Probing the imprint of dark mater particle properties on the matter power spectrum is one of the science goals of the Euclid mission \cite{Laureijs:2011gra}. The Euclid data will allow to constrain the dark matter-photon scattering through their impact on the linear matter power spectrum at considerably  smaller scales than the CMB  (c.f. Fig.~\ref{fig: pk}).

It is thus of interest for large scale surveys, such as Euclid \cite{Laureijs:2011gra}, to obtain precise theory predictions and up-to-date constraints on the possible interaction strength. To this end we reviewed the formalism governing the evolution of cosmological perturbations in the presence of dark matter-photon interactions, by explicitly deriving the modified Boltzmann equations, the source functions for the line of sight integration, and modifications to the tight coupling approximation. Various expressions have been used in the past for the tight coupling approximation. We show that these discrepancies only affect the computed CMB spectra at the $0.01~\upmu\mathrm{K}^2$ level, well below current experimental sensitivity. However, we find it valuable to have the formalism standing on a more solid footing.

Dark matter photon interactions lead to non-zero dark matter sound speed. We find that the inclusion of such a term affects the CMB predictions at a sub-percent level and hence can be neglected, while the effect on the matter power spectrum is much more important, as can be seen in Fig.~\ref{fig: soundspeed-pk}.

We perform a similar analysis to Ref.~\cite{Wilkinson:2013kia,DiValentino:2017oaw} but include the full 2015 Planck temperature, polarisation, and lensing data. Considering temperature data at high multipoles only, we already find that constraints on the interaction cross section tighten by 20\% with respect to previous results, giving $\sigma_{\mathrm{DM}-\gamma}\leq 2.25\times 10^{-6}\,\sigma_\mathrm{Th}\left(m_\mathrm{DM}/\mathrm{GeV}\right)$. Further improvement can be achieved by considering lensing and polarisation data, leading to an overall decrease by 35\% with respect to the results of Ref.~\cite{Wilkinson:2013kia}. However, our tightest constraint, $\sigma_{\mathrm{DM}-\gamma}\leq 1.49\times 10^{-6}\,\sigma_\mathrm{Th}\left(m_\mathrm{DM}/\mathrm{GeV}\right)$, has to be taken with a grain of salt as the Planck collaboration considers the high-multipole polarisation data as preliminary due to unresolved systematic effects of $\mathcal{O}(\upmu\mathrm{K}^2)$. We note that the imminent Planck data release is expected to provide more accurate high-multipole polarisation measurements, which might affect our $\ugc$ upper limit slightly.

\begin{figure}
\includegraphics{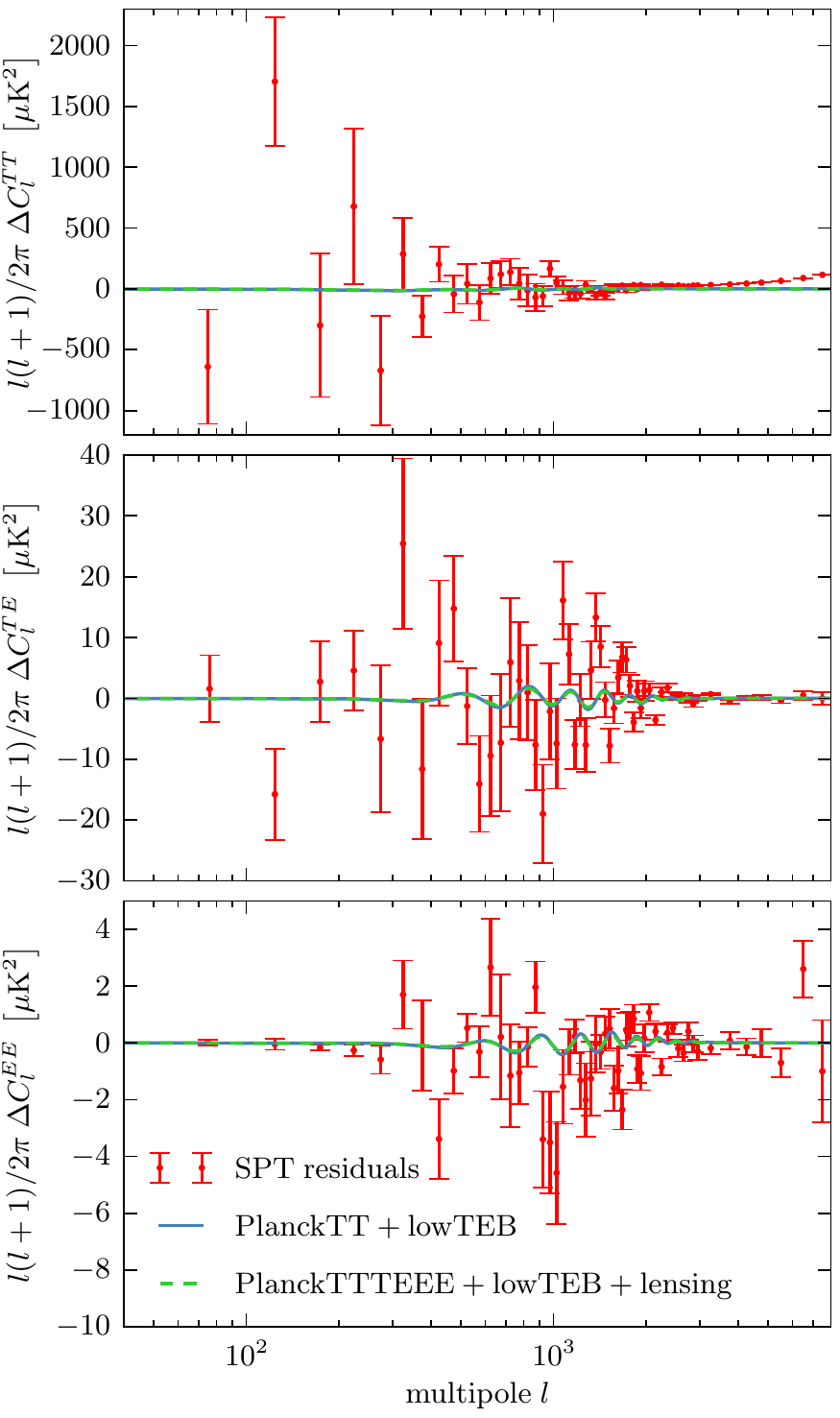}
\caption{Residuals between SPT measurements and the "Planck TT + lowTEB" $\Lambda$CDM best fit curve. Also shown is the impact, that the maximally allowed value of $\ugc$ (at 95\% confidence) has on the Planck best-fit $\Lambda$CDM spectrum for various datasets.}
\label{fig: residuals-spt}
\end{figure}

While preparing this manuscript, the SPTpol experiment released measurements of the EE and TE CMB polarisation, which constitute the most sensitive data at large multipoles, i.e. $l > 1050$ and $l > 1475$ respectively \cite{Henning:2017nuy}. This work also includes the determination of the temperature auto-correlation (TT) spectrum up to multipoles as large as $l=8000$. However, significant atmospheric contaminations where left in the TT spectrum so this dataset was not used to constrain cosmology in Ref.~\cite{Henning:2017nuy}.

In Fig.~\ref{fig: residuals-spt} we show the residuals between the ``Planck TT + lowTEB" $\Lambda$CDM best fit model and the SPTpol data. This figure also depicts the impact, which the maximally allowed value of $\ugc$ (at 95\% confidence) would have on the best-fit $\Lambda$CDM model for different datasets. To obtain these curves we first compute the CMB spectra from the best fit values found by the Planck collaboration for the six-parameter $\Lambda$CDM model and the specific dataset. We then add the maximally allowed value of $\ugc$ for the specific dataset (cf. Tab~\ref{tab: PlanckTT+lowTEB} and Tab.~\ref{tab: PlanckTTTEEE+lowTEB}) but keep the value of the six $\Lambda$CDM parameters the same as determined by the Planck collaboration in absence of interactions, and finally we compute the difference between the scenario with and without the dark matter-photon interactions.

From Fig.~\ref{fig: residuals-spt} it is obvious that dark matter-photon scattering has the most sizeable effect on the EE and TE spectra at intermediate multipoles, up to $l\simeq2000$. Furthermore, we see that these interactions are already tightly constrained by the Planck TT spectrum. Indeed, the effect that the maximally allowed values of $\ugc$ have on the TE and EE spectrum is smaller than the SPTpol $1\sigma$ errorbars at all scales.

Therefore, while the new SPTpol data has not been included in our runs, it is very unlikely that its inclusion would make any difference to the upper bound on $\ugc$. Adding the SPTpol data might narrow the allowed regions for the $\Lambda$CDM parameters, but we do not observe any parameter degeneracies with $\ugc$ that would suggest this could have a huge impact on our result.

\section*{Acknowledgement}
The authors thank Eleonora Di Valentino, Miguel Escudero and Ryan Wilkinson for useful discussion. This project has received funding/support from the European Union’s Horizon 2020 research and innovation program under the Marie Sklodowska-Curie grant agreement No 674896. This research was supported in part by Perimeter Institute for Theoretical Physics. Research at Perimeter Institute is supported by the Government of Canada through Industry Canada and by the Province of Ontario through the Ministry of Economic Development and Innovation. 

\bibliography{gcdm-CMB-paper}
\end{document}